\documentclass[aps,twocolumn,preprintnumbers,floats]{revtex4}\def\@cite#1#2{\textsuperscript{[{#1\if@tempswa , #2\fi}]}}
\usepackage{float}
\usepackage{ulem}
\usepackage{graphicx}
\usepackage{amsmath}
\usepackage{bm}
\usepackage{amsfonts}
\usepackage{amssymb}
\usepackage{color}
\usepackage{xcolor}
\usepackage{subfigure}
\usepackage{epsfig}
\usepackage{morefloats}
\usepackage{multirow}
\usepackage{graphicx,booktabs}
\usepackage{mathrsfs}
\usepackage{txfonts}
\usepackage{pifont}
\usepackage{indentfirst}
\usepackage{graphicx,booktabs}
\usepackage{longtable,lscape}
\usepackage[figuresright]{rotating}
\usepackage[colorlinks, citecolor=blue,anchorcolor=red,menucolor=red, linkcolor=red,filecolor=red,urlcolor=blue,frenchlinks=red]{hyperref}

\begin{document}
	
\title{All-heavy pentaquarks }
\author{Zhi-Biao Liang$^{1}$, Feng-Xiao Liu$^{2,3}$~\footnote{E-mail: liufx@ihep.ac.cn},
and
Xian-Hui Zhong$^{1,3}$~\footnote{E-mail: zhongxh@hunnu.edu.cn}}
\affiliation{ 1) Department of Physics, Hunan Normal University, and Key Laboratory of Low-Dimensional Quantum Structures and Quantum Control of Ministry of Education, Changsha 410081, China}
\affiliation{ 2) Institute of High Energy Physics, Chinese Academy of Sciences, Beijing 100049, China}
\affiliation{ 3) Synergetic Innovation Center for Quantum Effects and Applications (SICQEA), Hunan Normal University, Changsha 410081, China}

\begin{abstract}
	
In a nonrelativistic potential quark model framework, we carry out a calculation
of the mass spectrum for the low-lying $1S$ all-heavy pentaquark state by adopting the explicitly correlated Gaussian method.
The obtained states are compact and lie far above
the lowest dissociation baryon-meson threshold. Moreover,
using the obtained masses and wave functions we evaluate the fall-apart decay properties within a
quark-exchange model. The results show that the $1S$ all-heavy pentaquark states
have a fairly narrow fall-apart width, which scatters in the range of $\sim0.1-4.0$ MeV.
Their dominant fall-apart decay channels may be ideal for searching for their signals in future experiments.

\end{abstract}


\maketitle

\section{Introduction}

Searching for genuine exotic multiquark states beyond the conventional
meson ($q\bar{q}$) and baryon ($qqq$) states has been one of the most important
initiatives since the establishment of quark model in 1964~\cite{Gell-Mann:1964ewy,Zweig:1964ruk,Zweig:1964jf}.
Since the discovery of $X(3872)$ by Belle in 2003~\cite{Belle:2003nnu},
many tetraquark candidates, such as the series hidden-charmed/bottom $XYZ$ states~\cite{ParticleDataGroup:2022pth},
the doubly-charmed state $T_{cc}(3875)^+$~\cite{LHCb:2021auc,LHCb:2021vvq}, and charmed-strange
states~\cite{LHCb:2022sfr,LHCb:2020bls}, have been observed in experiments.
Furthermore, the exotic $P_c$~\cite{LHCb:2015yax,LHCb:2016ztz,LHCb:2019kea} and $P_{cs}$~\cite{LHCb:2022ogu} states as candidates of pentaquark
states were also reported by the LHCb collaboration.
All of these observed exotic states contain two or three light quarks.
They may be hadronic molecular states usually arising from the one-light-meson exchanges~\cite{Dong:2021bvy,Guo:2017jvc,Liu:2019zoy,Chen:2016qju}, or genuine compact multiquark states arising from the
one-gluon exchange (OGE)~\cite{Liu:2019zoy,Chen:2016qju}.
There are many debates about whether these exotic states are hadronic molecular states or genuine multiquark states in the literature.
This dilemma may be largely alleviated for the all-heavy
multiquarks, which are most likely to form genuine multiquark states since the one-light-meson exchanges are absent.
It should be emphasized that one cannot exclude the possibility that some molecular states
may be still formed by other dynamical mechanisms~\cite{Brambilla:2015rqa,Dong:2022rwr,Dong:2021lkh,Liu:2023gla,Liu:2024pio,Gong:2020bmg,Liu:2021pdu,Lyu:2021qsh,Mathur:2022ovu}.
As a whole, the study of the all-heavy multiquarks may provide an interesting way for establishing
genuine multiquark states, or revealing some special dynamical mechanisms if they form molecular states.

Impressively, some all-heavy multiquark states have been observed in LHC experiments.
In 2020, the LHCb Collaboration observed a narrow structure $X(6900)$ together a
broad structure ranging from 6.2 to 6.8 GeV in the di-$J/\psi$ invariant mass spectrum~\cite{LHCexp}.
Later in 2022, the $X(6900)$ was confirmed in the same final state by both the ATLAS~\cite{ATLASexp} and CMS~\cite{CMSexp} collaborations.
Moreover, in the lower mass region the CMS measurements show that a clear resonance $X(6600)$ lies in the di-$J/\psi$ spectrum.
These clear structures may be evidence for genuine all-charmed tetraquark $cc\bar{c}\bar{c}$ states~\cite{Chao:2020dml,liu:2020eha,Liu:2021rtn,Lu:2020cns,Wang:2021kfv,Wu:2024euj,Badalian:2023rpd,Chen:2022asf,
Bedolla:2019zwg,Zhao:2020jvl,Yu:2022lak,Zhang:2022qtp}.
The discovery of $cc\bar{c}\bar{c}$ states has demonstrated the powerful abilities of LHC in production of
fully-heavy hadrons, and also indicates that the all-heavy pentaquark should exist.
Thus, one may expect to observe some all-heavy pentaquarks in forthcoming experiments.
Stimulated by these, in the recent years some studies of the
mass spectrum of the all-heavy pentaquarks have been carried out within some models,
such as the QCD sum rules~\cite{Zhang:2020vpz,Wang:2021xao,Azizi:2024ito}, chromomagnetic interaction (CMI) model~\cite{An:2020jix},
MIT bag model~\cite{Zhang:2023hmg}, effective mass and screened charge model~\cite{Rashmi:2024ako},
and various potential models with different numerical methods and approximations~\cite{An:2022fvs,Yan:2021glh,Yan:2023kqf,Yang:2022bfu,Gordillo:2023tnz,Gordillo:2024blx}.
It should be mentioned that there are strong model dependencies in the predictions.
For example, for the $1S$-wave $cccc\bar{c}$ states the predicted masses
scatter in a very large range of $\sim7.4-8.5$ GeV.

In this work, we carry out a dynamical calculation of the mass spectra of
the $1S$-wave all-heavy pentaquarks with a nonrelativistic quark potential model.
This model is based on the Hamiltonian of the Cornell model~\cite{Eichten:1978tg},
whose parameters have been well determined based
on the successful description of the heavy quarkonium spectra
~\cite{Deng:2016stx,Li:2019tbn,Li:2019qsg}, and triply charmed and bottom baryon spectra~\cite{Liu:2019vtx}.
This model has been extended to study the mass spectra of all-heavy tetraquarks
by our group in Refs.~\cite{Liu:2019zuc,Liu:2021rtn,liu:2020eha}.
To solve the five-body problem accurately, we adopt the explicitly correlated Gaussian (ECG)
method~\cite{Mitroy:2013eom,Varga:1995dm},
in which a variational trial spatial wave function is expanded with the most
general nondiagonal Gaussian basis functions associated to the Jacobi coordinates.
This is one of the most powerful approaches currently used for calculating the properties of few-body
systems, and offers great flexibility, and high accuracy~\cite{Mitroy:2013eom}.
Based on the ECG numerical method, recently the mass spectrum of charmed-strange
tetraquarks was calculated by our group~\cite{Liu:2022hbk}.

Moreover, considering the fact that all of the states obtained in the present work lie far above
their lowest dissociation baryon-meson threshold, we further evaluate the fall-apart
decays of the $1S$-wave pentaquark states in a
quark-exchange model~\cite{Barnes:2000hu,Barnes:1991em}, with which
a good description of the low-energy $S$-wave phase shift for the $I=2$ $\pi\pi$
scattering at the quark level has been achieved.
Recently, this phenomenological model has been widely adopted to study
the fall-apart decay properties of the
multi-quark states in the literature~\cite{Yang:2021sue,Wang:2019spc,Xiao:2019spy,Wang:2020prk,Han:2022fup,Liu:2024fnh,
Liu:2022hbk,liu:2020eha,Wang:2021hql,Liu:2014eka}.
It should be mentioned that the study of the fall-apart decay properties of the
all-heavy pentaquarks is scare, only a few discussions based on the wave function overlapping
between the initial and final states can be found in Ref.~\cite{An:2020jix}.


As follows, firstly we give a brief introduction to our
framework in Sec.~\ref{Fram}. Then we give our numerical results and discussions for
the $S$-wave states of all-heavy pentaquarks in Sec.~\ref{RESULTS AND DISCUSSIONS}.
Finally, a summary is given in Sec.~\ref{sum}.

\section{Framework}\label{Fram}

\subsection{Spectrum}

In this section, we first give a brief review of the nonrelativistic quark potential
model (NRQPM). Then, we introduce the quark model classification
of the pentaquarks based on the symmetries. Finally, we give a brief
introduction to the ECG method what we adopt in the calculations.
	
\begin{table}[hptb]
	\caption{The quark model parameters determined by fitting the
meson mass spectra. The unit of the meson masses is MeV. }\label{Fitting results}
	\begin{tabular}{cccccc}
		\hline\hline
\multicolumn{6}{c}{Quark model parameters}\\ \hline
\multicolumn{3}{c}{$m_c/m_b$(GeV)}&\multicolumn{3}{c}{1.483/4.852}\\
\multicolumn{3}{c}{$\alpha_{cc}/\alpha_{bc}/\alpha_{bb}$}&\multicolumn{3}{c}{0.5461/0.5021/0.4311}\\
\multicolumn{3}{c}{$\sigma_{cc}/\sigma_{bc}/\sigma_{bb}$(GeV) }&\multicolumn{3}{c}{1.1384/1.3000/2.3200}\\
\multicolumn{3}{c}{$b$(GeV$^2$)}&\multicolumn{3}{c}{0.1425}\\ \hline\hline
\multicolumn{6}{c}{Mass spectrum}\\
\hline
 Meson      & Theor. & Exp.~\cite{ParticleDataGroup:2022pth} &         Meson         & Theor. & Exp.~\cite{ParticleDataGroup:2022pth} \\
\hline
		 $J/\psi(1S)$   & 3097 & 3097  &   $B_{c}(2S)^{\pm}$   & 6871  & 6871  \\
		$\eta_{c}(1S)$  & 2983 & 2984  & $\Upsilon\text{(1S)}$ & 9460  & 9460  \\
		  $\psi(2S)$    & 3679 & 3686  &    $\eta_{b}(1S)$     & 9390  & 9399  \\
		$\eta_{c}(2S)$  & 3635 & 3638  &    $\Upsilon(2S)$     & 10024 & 10023 \\
		$\chi_{c0}(1P)$ & 3415 & 3415  &    $\eta_{b}(2S)$     & 10005 & 9999  \\
		$\chi_{c1}(1P)$ & 3516 & 3511  &      $h_{b}(1P)$      & 9941  & 9899  \\
		$\chi_{c2}(1P)$ & 3552 & 3556  &    $\chi_{b0}(1P)$    & 9859  & 9859  \\
		  $h_{c}(1P)$   & 3522 & 3525  &    $\chi_{b1}(1P)$    & 9933  & 9893  \\
		  $B_{c}^{+}$   & 6271 & 6274  &    $\chi_{b2}(1P)$    & 9957  & 9912  \\ \hline\hline
	\end{tabular}
\end{table}

\subsubsection{Hamiltonian}

In NRQPM, the Hamiltonian for a pentaquark system is given by
\begin{eqnarray}
	H=\sum^{5}_{i}(m_{i}+T_{i})-T_{cm}+\sum^{5}_{i<j}V_{ij}(r_{ij}),
\end{eqnarray}
where $m_{i}$ and $T_{i}$ represent the mass and kinetic energy of the $i$-th quark, respectively;
$T_{cm}$ represents the kinetic energy of the center
of mass of the pentaquark system; $r_{ij}=|\boldsymbol{r_i}-\boldsymbol{r_j}|$ is the distance between the $i$-th and
$j$-th quarks; while $V_{ij}$ is an effective potential between them. In this work, the effective potential $V_{ij}$ adopts a widely used form,
\begin{eqnarray}\label{vij}
	\begin{aligned}		V_{ij}(r_{ij})&=-\frac{3}{16}\left(\boldsymbol{\lambda}_{i}\cdot\boldsymbol{\lambda}_{j}\right)\left(b r_{ij}-\frac{4}{3}\frac{\alpha_{ij}}{r_{ij}}\right)\\		 &-\frac{\alpha_{ij}}{4}\left(\boldsymbol{\lambda}_i\cdot\boldsymbol{\lambda}_j\right)\left[\frac{\pi}{2}\cdot\frac{\sigma_{ij}^3e^{-\sigma_{ij}^2r_{ij}^2}}{\pi^{3/2}}\cdot\frac{16}{3m_im_j}
		\left(\textbf{S}_i\cdot\textbf{S}_j\right)\right ],
	\end{aligned}
\end{eqnarray}
where $\boldsymbol{\lambda}_{i,j}$ are the color operators acting on the $i,j$-th quarks,
$\boldsymbol{S}_{i,j}$ represent the spin operators of the $i,j$-th quarks. The parameters
$b$ and $\alpha_{ij}$ denote the strength of the confinement and
strong coupling of the OGE potential, respectively.
In this work, the parameter set \{$b$, $\alpha_{ij}$, $\sigma_{ij}$, $m_i$ \} is given in
Table~\ref{Fitting results}, which is taken the same as that
in our previous works~\cite{Liu:2019zuc,Liu:2019vtx}. These parameters were determined through
the fitting of the mass spectra of charmonium, bottomonium, and $B_c$ mesons.
The theoretical masses compared with the data are collected in
Table~\ref{Fitting results} as well. Finally, it should be mentioned that in this work we only consider the
low-lying $1S$-wave pentaquark states without any orbital excitations, thus,
the spin-orbit and tensor potentials are not included.

\subsubsection{Pentaquark configuration}

The pentaquark configuration can be expressed as a product of the flavor, spatial, spin, and color parts, i.e.,
\begin{eqnarray}
	|\Psi_{5q}\rangle=|flavor\rangle \otimes |spatial \rangle \otimes |spin \rangle \otimes |color \rangle.
\end{eqnarray}
In the flavor space, the available configurations for all-heavy pentaquarks are ``$cccc\bar{c}$, $cccc\bar{b}$, $bbbb\bar{b}$, $bbbb\bar{c}$,
$cccb\bar{b}$, $cccb\bar{c}$, $bbbc\bar{c}$, $bbbc\bar{b}$, $ccbb\bar{c}$, $bbcc\bar{b}$''.
The spin and color parts can be constructed according to the permutation symmetry.
For the low-lying $1S$ pentaquark states without any excitations,
their spatial wave functions are symmetrical when exchanging the
coordinates of any two quarks. To obtain the spatial part of the pentaquarks, one should
solve the Schr\"{o}dinger equation.

Due to the pentaquark system containing identical quarks, it must satisfy the Pauli principle.
Considering the permutation symmetry of identical quarks, the pentaquark
configurations can be classified into three categories, which are denoted by
\begin{eqnarray}
	\begin{aligned}
		\{1234\}\bar{5}     & : \{cccc\}\bar{c}, \{cccc\}\bar{b}, \{bbbb\}\bar{b}, \{bbbb\}\bar{c}, \\
		\{123\}4\bar{5}     & : \{ccc\}b\bar{b}, \{ccc\}b\bar{c}, \{bbb\}c\bar{c}, \{bbb\}c\bar{b}, \\
		\{12\}\{34\}\bar{5} & : \{cc\}\{bb\}\bar{c}, \{bb\}\{cc\}\bar{b}.
	\end{aligned}
\end{eqnarray}
The quarks in $\{\}$ should satisfy the requirements of the permutation symmetry.

For a pentaquark system, one can construct three colorless configurations (color singlets) in the color space based on
the SU(3)-group representation theory. Their representations with Young
tableaux are given by~\cite{Park:2017jbn}
	\begin{eqnarray}\label{color}
		\begin{aligned}
			C_1=
			\begin{small}
				\begin{tabular}{ll}
					\cline{1-2}
					\multicolumn{1}{|l|}{1} & \multicolumn{1}{|l|}{4} \\ \cline{1-2}
					\multicolumn{1}{|l|}{2} &                         \\ \cline{1-1}
					\multicolumn{1}{|l|}{3} &                         \\ \cline{1-1}
				\end{tabular}
			\end{small}_3
			\otimes(5)_{\bar{3}},\  \
		\end{aligned}
		\begin{aligned}
			C_2=\begin{small}
				\begin{tabular}{ll}
					\cline{1-2}
					\multicolumn{1}{|l|}{1} & \multicolumn{1}{|l|}{2} \\ \cline{1-2}
					\multicolumn{1}{|l|}{3} &                         \\ \cline{1-1}
					\multicolumn{1}{|l|}{4} &                         \\ \cline{1-1}
				\end{tabular}
			\end{small}_3\otimes(5)_{\bar{3}}, \ \
		\end{aligned}
	\begin{aligned}
		C_3=\begin{small}
			\begin{tabular}{ll}
			\cline{1-2}
			\multicolumn{1}{|l|}{1} & \multicolumn{1}{|l|}{3} \\ \cline{1-2}
			\multicolumn{1}{|l|}{2} &                         \\ \cline{1-1}
			\multicolumn{1}{|l|}{4} &                         \\ \cline{1-1}
		\end{tabular}
		\end{small}_3\otimes(5)_{\bar{3}}.
	\end{aligned}
\end{eqnarray}
By using the C-G coefficients of SU(3) group~\cite{Kaeding:1995vq}, one can explicitly write out
the three color configurations, which can be found in the literature, such as Ref.~\cite{Zhang:2023hmg}.
The three colorless configurations $C_1$, $C_2$, and $C_3$ can be also expressed with the form of
baryon-meson structures, i.e.,
\begin{eqnarray}
	\label{eq color}
	C_1=\begin{tabular}{|c|}
		\hline
		1 \\ \hline
		2 \\ \hline
		3 \\ \hline
	\end{tabular}_1 (4\bar{5})_1, \quad C_2=\begin{tabular}{|c|c|}
	\hline
	1 & 2\\
	\hline
	3 & \multicolumn{1}{c}{}\\
	\cline{1-1}
	\end{tabular}_8\otimes (4\bar{5})_8, \quad C_3=\begin{tabular}{|c|c|}
	\hline
	1 & 3\\
	\hline
	2 & \multicolumn{1}{c}{}\\
	\cline{1-1}
	\end{tabular}_8\otimes (4\bar{5})_8.
\end{eqnarray}
It is found that the $C_1$ configuration is a simple combination of the color singlets of baryon and meson.

For a pentaquark system, the total spin quantum numbers
are possibly $J=5/2,3/2,1/2$. Based on the SU(2) symmetry, one can construct
the configurations with $J=5/2,3/2,1/2$ in the spin space. Their representations with Young
tableaux are given by
\begin{eqnarray}\label{a82}
	J=\frac{5}{2}: S_1=\begin{tabular}{lllll}
		\cline{1-4}
		\multicolumn{1}{|l|}{1} & \multicolumn{1}{|l|}{2} & \multicolumn{1}{|l|}{3} & \multicolumn{1}{|l|}{4} & 5. ~~~~~~~~~~~~~~~~~~~~~~~~~~~~~~~~~~~~~~~~\\ \cline{1-4}
\end{tabular}
\end{eqnarray}
\begin{eqnarray}\label{a32}
	\begin{aligned}
		\smallskip\smallskip
		J=\frac{3}{2}:S_2&=\begin{tabular}{llll}
			\cline{1-4}
			\multicolumn{1}{|l|}{1} & \multicolumn{1}{|l|}{2} & \multicolumn{1}{|l|}{3} & \multicolumn{1}{|l|}{4} \\ \cline{1-4}
			5                       &                         &                         &\\
		\end{tabular}, \quad\quad
		S_3=\begin{tabular}{llll}
			\cline{1-3}
			\multicolumn{1}{|l|}{1} & \multicolumn{1}{|l|}{2} & \multicolumn{1}{|l|}{3} & 5 \\ \cline{1-3}
			\multicolumn{1}{|l|}{4} &                         &                         &   \\ \cline{1-1}
		\end{tabular},\\\smallskip\smallskip
		S_4&=\begin{tabular}{llll}
			\cline{1-3}
			\multicolumn{1}{|l|}{1} & \multicolumn{1}{|l|}{3} & \multicolumn{1}{|l|}{4} & 5 \\ \cline{1-3}
			\multicolumn{1}{|l|}{2} &                         &                         &   \\ \cline{1-1}
		\end{tabular}, \quad\quad
		S_5=\begin{tabular}{llll}
			\cline{1-3}
			\multicolumn{1}{|l|}{1} & \multicolumn{1}{|l|}{2} & \multicolumn{1}{|l|}{4} & 5 \\ \cline{1-3}
			\multicolumn{1}{|l|}{3} &                         &                         &   \\ \cline{1-1}
		\end{tabular}.~~~~~~~~~~~~~
	\end{aligned}
\end{eqnarray}
\begin{eqnarray}\label{a12}
	\begin{aligned}
		\smallskip\smallskip
		J=\frac{1}{2}:S_6&=\begin{tabular}{lll}
			\cline{1-3}
			\multicolumn{1}{|l|}{1} & \multicolumn{1}{|l|}{2} & \multicolumn{1}{|l|}{3} \\ \cline{1-3}
			\multicolumn{1}{|l|}{4} & 5                       &                         \\ \cline{1-1}
		\end{tabular}, \quad
		S_7=\begin{tabular}{lll}
			\cline{1-3}
			\multicolumn{1}{|l|}{1} & \multicolumn{1}{|l|}{3} & \multicolumn{1}{|l|}{4} \\ \cline{1-3}
			\multicolumn{1}{|l|}{2} & 5                       &                         \\ \cline{1-1}
		\end{tabular}, \quad
		S_8=\begin{tabular}{lll}
			\cline{1-3}
			\multicolumn{1}{|l|}{1} & \multicolumn{1}{|l|}{2} & \multicolumn{1}{|l|}{4} \\ \cline{1-3}
			\multicolumn{1}{|l|}{3} & 5                       &                         \\ \cline{1-1}
		\end{tabular}, \\\smallskip\smallskip
		S_9&=\begin{tabular}{lll}
			\cline{1-2}
			\multicolumn{1}{|l|}{1} & \multicolumn{1}{|l|}{2} & 5 \\ \cline{1-2}
			\multicolumn{1}{|l|}{3} & \multicolumn{1}{|l|}{4} &   \\ \cline{1-2}
		\end{tabular}, \quad
		S_{10}=\begin{tabular}{lll}
			\cline{1-2}
			\multicolumn{1}{|l|}{1} & \multicolumn{1}{|l|}{3} & 5 \\ \cline{1-2}
			\multicolumn{1}{|l|}{2} & \multicolumn{1}{|l|}{4} &   \\ \cline{1-2}
		\end{tabular}.
	\end{aligned}
\end{eqnarray}
Combining the C-G coefficient of SU(2) group, one can obtain the spin wave function
$\psi_{S_i}|JJ_z\rangle$ corresponding to a special Young tableau.

With the configurations of the color and spin spaces,
one can further construct the configurations of the spin$\otimes$color space.
For the low-lying $1S$-wave pentaquark configurations, both the spatial and flavor parts are symmetric when
one exchanges any two identical quarks, thus, the spin$\otimes$color part should be antisymmetric.
To get the coupling configurations in the spin$\otimes$color space, we need the C-G coefficients of permutation group,
which are taken from Ref.~\cite{Stancu:1999qr}.

For a pentaquark system with $\{1234\}\bar{5}$ symmetry, there are two configurations with $J^P=\frac{3}{2}^-$ and $\frac{1}{2}^-$.
By combining the C-G coefficients of the $S_4$ group, one can obtain the color-spin wave functions,
\begin{eqnarray}
		1S_{\frac{3}{2}^-}(\{1234\}\bar{5}) & =\frac{1}{\sqrt{3}}\left(C_1 S_3+C_2 S_4-C_3 S_5\right), \label{1234} \\
		1S_{\frac{1}{2}^-}(\{1234\}\bar{5}) & =\frac{1}{\sqrt{3}}\left(C_1 S_6+C_2 S_7-C_3 S_8\right)\label{1234b}.
\end{eqnarray}
It should be mentioned that there is no $J^P=\frac{5}{2}^-$ configurations. In this case,
the $\{1234\}\bar{5}$ symmetry requires the color configurations are fully antisymmetric
since the spin configurations are fully symmetric, however, the color configurations given in Eq.~(\ref{color})
cannot satisfy this requirement.

For a pentaquark system with $\{123\}4\bar{5}$ symmetry, by combining the C-G coefficients of the $S_3$ group,
in the spin$\otimes$color space one can get one
$J^P=\frac{5}{2}^-$ configuration,
\begin{eqnarray}\label{equation 25}
	\begin{aligned}
		1S_{\frac{5}{2}^-}(\{123\}4\bar{5})   & = C_1 S_1,
	\end{aligned}
\end{eqnarray}
three $J^P=\frac{3}{2}^-$ configurations,
\begin{eqnarray}
		1S_{\frac{3}{2}^-}(\{123\}4\bar{5})_1  &=& C_1 S_2,   \\
		1S_{\frac{3}{2}^-}(\{123\}4\bar{5})_2  &=& C_1 S_3, \label{equation 123a} \\
		1S_{\frac{3}{2}^-}(\{123\}4\bar{5})_3  &=& \sqrt{\frac{1}{2}}(C_3 S_5-C_2 S_4)\label{equation 123b},
\end{eqnarray}
and three $J^P=\frac{1}{2}^-$ configurations
\begin{eqnarray}
		1S_{\frac{1}{2}^-}(\{123\}4\bar{5})_1 & =&C_1 S_6,   \\
		1S_{\frac{1}{2}^-}(\{123\}4\bar{5})_2 & =& \sqrt{\frac{1}{2}}(C_3 S_8-C_2 S_7), \\
		1S_{\frac{1}{2}^-}(\{123\}4\bar{5})_3 & =& \sqrt{\frac{1}{2}}(C_3 S_9-C_2 S_{10}).
\end{eqnarray}

For a pentaquark system with $\{12\}\{34\}\bar{5}$ symmetry, there is a $J^P=\frac{5}{2}^-$ configuration,
\begin{eqnarray}
	\begin{aligned}\label{eq 2c5/2}
		1S_{\frac{5}{2}^-}(\{12\}\{34\}\bar{5})   & =  \sqrt{\frac{1}{3}}C_1 S_1-\sqrt{\frac{2}{3}}C_3 S_1,
	\end{aligned}
\end{eqnarray}
four $J^P=\frac{3}{2}^-$ configurations,
\begin{eqnarray}\label{eq 2c3/2}
		1S_{\frac{3}{2}^-}(\{12\}\{34\}\bar{5})_1 &=  & \sqrt{\frac{1}{3}}C_1 S_2-\sqrt{\frac{2}{3}}C_3 S_2, \\
		1S_{\frac{3}{2}^-}(\{12\}\{34\}\bar{5})_2  &= & \frac{1}{3}(C_1 S_3-2C_3 S_5)                \nonumber          \\
		                                           &  & +\frac{\sqrt{2}}{3}(C_1 S_5-C_3 S_3),                \\
		1S_{\frac{3}{2}^-}(\{12\}\{34\}\bar{5})_3  &= & C_2 S_4                                             \\
		1S_{\frac{3}{2}^-}(\{12\}\{34\}\bar{5})_4  &= & \frac{1}{3}(C_3 S_5-2C_1 S_3)           \nonumber              \\
		                                           &  & +\frac{\sqrt{2}}{3}(C_1 S_5-C_3 S_3),
\end{eqnarray}
and four $J^P=\frac{1}{2}^-$ configurations,
\begin{eqnarray}\label{eq 2c1/2}
		1S_{\frac{1}{2}^-}(\{12\}\{34\}\bar{5})_1  &= & \frac{1}{3}(C_1 S_6-2C_3 S_8)               \nonumber \\
		                                           &  & +\frac{\sqrt{2}}{3}(C_1 S_8-C_3 S_6),                \\
		1S_{\frac{1}{2}^-}(\{12\}\{34\}\bar{5})_2  &= & C_2 S_7,                                            \\
		1S_{\frac{1}{2}^-}(\{12\}\{34\}\bar{5})_3  &= & \frac{1}{3}(C_3 S_8-2C_1 S_6)               \nonumber          \\
		                                           &  & +\frac{\sqrt{2}}{3}(C_1 S_8-C_3 S_6),                \\
		1S_{\frac{1}{2}^-}(\{12\}\{34\}\bar{5})_4  &= & \sqrt{\frac{1}{3}}C_1 S_9-\sqrt{\frac{2}{3}}C_3 S_9.
\end{eqnarray}

Finally, it should be mentioned that some configurations with $J^P=\frac{1}{2}^-$ and $\frac{3}{2}^-$ for
the pentaquark systems with $\{123\}4\bar{5}$ and $\{12\}\{34\}\bar{5}$ symmetries
are different from those constructed by using the $S_4$ group in the literature~\cite{An:2020jix,An:2022fvs,Zhang:2023hmg}.
For example, if constructing the configurations with the $S_4$ group for the system with the $\{123\}4\bar{5}$ symmetry, one has two
$J^P=\frac{3}{2}^-$ configurations, $\Phi_1=(C_1S_3+C_2S_4-C_3S_5)/\sqrt{3}$
and $\Phi_2=[2C_1S_3-(C_2S_4-C_3S_5)]/\sqrt{6}$. It is found that
these representations of $S_4$ Group should be further broken into those of $S_3$ Group
given in Eqs.~(\ref{equation 123a}) and (\ref{equation 123b}), because the interactions for the $C_1S_3$ basis 
are very different from those for the $C_2S_4$ and $C_3S_5$. To know more technical details for constructing the pentaquark configurations
in the spin$\otimes$color space, one can refer to the Appendix.
Where, three examples for constructing the spin-color configurations with
the $\{1234\}\bar{5}$, $\{123\}4\bar{5}$, $\{12\}\{34\}\bar{5}$ symmetries, have been given, respectively.

\subsubsection{Numerical method}

To solve the five-body problem accurately, we adopt the ECG method~\cite{Mitroy:2013eom,Varga:1995dm}.
It is a well-established variational method to solve quantum few-body problems
in molecular, atomic, and nuclear physics. The spatial part of the wave function
for a pentaquark system without any angular momenta is expanded in terms of ECG
basis set. Such a basis function can be expressed as
\begin{eqnarray}	 \psi(\boldsymbol{r}_{1},\boldsymbol{r}_{2},\boldsymbol{r}_{3},\boldsymbol{r}_{4},\boldsymbol{r}_{5})=\exp\left(-\sum\limits_{i<j}^5a_{ij}\boldsymbol{r}_{ij}^2\right),
\end{eqnarray}
where $a_{ij}$ are variational parameters. For a pentaquark system with $\{1234\}\bar{5}$ symmetry,
one has two independent variational parameters, i.e., $a_{12}=a_{13}=a_{14}=a_{23}=a_{24}=a_{34}=a$ and $a_{15}=a_{25}=a_{35}=a_{45}=b$,
due to the symmetry of identical quarks. Similarly, for a pentaquark system with $\{123\}4\bar{5}$-symmetry, there are four variational parameters, $a_{12}=a_{13}=a_{23}=a$, $a_{14}=a_{24}=a_{34}=b$, $a_{15}=a_{25}=a_{35}=c$ and $a_{45}=d$.
While for a pentaquark system with $\{12\}\{34\}\bar{5}$ symmetry, there are also five independent variational parameters, $a_{12}=a$, $a_{13}=a_{14}=a_{23}=a_{24}=b$, $a_{15}=a_{25}=c$,
$a_{34}=d$ and $a_{35}=a_{45}=f$.

It is convenient to use a set of the Jacobi
coordinates $\boldsymbol{\xi}=\left(\boldsymbol{\xi}_1, \boldsymbol{\xi}_2, \boldsymbol{\xi}_3,
\boldsymbol{\xi}_4\right)$ instead of position vectors $\boldsymbol{r_{i}}$ ($i=1-5$).
For example, one can take a set of Jacobi coordinates as follows,
\begin{eqnarray}
	\begin{aligned}
		\boldsymbol{\xi_1} & = \boldsymbol{r_1}-\boldsymbol{r_2},                                                                                                        \\
		\boldsymbol{\xi_2} & = \boldsymbol{r_3}-\frac{m_1\boldsymbol{r_1}+m_2\boldsymbol{r_2}}{m_1+m_2},                                                                 \\
		\boldsymbol{\xi_3} & = \boldsymbol{r_4}-\boldsymbol{r_5},                                                                                                        \\
		\boldsymbol{\xi_4} & = \frac{m_1\boldsymbol{r_1}+m_2\boldsymbol{r_2}+m_3\boldsymbol{r_3}}{m_1+m_2+m_3}-\frac{m_4\boldsymbol{r_4}+m_5\boldsymbol{r_5}}{m_4+m_5}.
	\end{aligned}
\end{eqnarray}
Then, the correlated Gaussian basis function $\psi$ can be rewritten as
\begin{eqnarray}
	G(\boldsymbol{\xi},A)=\exp(-\boldsymbol{\xi}^TA\boldsymbol{\xi}),
\end{eqnarray}
where $A$ is a $4\times4$ matrix, which is related to the variational parameters.
The spatial part of the trial wave function can be expanded with a set of correlated Gaussians:
\begin{eqnarray}\label{spatial}
	\Psi(\boldsymbol{\xi},A)=\sum\limits_{k=1}^N C_k G(\boldsymbol{\xi},A_k),
\end{eqnarray}
where $N$ is the number of Gaussian basis functions. The accuracy of the trial function depends on
the number $N$ and the nonlinear parameter matrix $A_k$. In our
calculations, following the method of Ref.~\cite{Hiyama:2003cu}, we let the
variational parameters form a geometric progression. For example,
for a variational parameter $a$, we take
\begin{eqnarray}
	a_i=\frac{1}{2\left(a_1q^{i-1}\right)^2}\quad\left(i=1,\cdots,n_{max}\right).
\end{eqnarray}
The Gaussian size parameters $\{a_1,q,n_{max}\}$ will be determined through the variation method.
In the calculations, the final results should be stable and independent with these parameters.

For a given pentaquark configuration, one can work out the Hamiltonian matrix elements,
\begin{eqnarray}
	H_{kk'}=\langle \psi_{CS}G(\boldsymbol{\xi},A_k) | H |\psi_{CS}G(\boldsymbol{\xi},A_{k'}) \rangle,
\end{eqnarray}
where $\psi_{CS}$ is the spin-color wave function.
Then, by solving the generalized matrix eigenvalue problem,
\begin{eqnarray}
	\sum_{k'=1}^{N}(H_{kk'}-EN_{kk'})C_{k'}=0,
\end{eqnarray}
one can obtain the eigenenergy $E$, and the expansion coefficients
$\{C_k\}$. The $N_{kk'}$ is an overlap factor defined by $N_{kk'}=\langle G(\boldsymbol{\xi},A_k)|G(\boldsymbol{\xi},A_{k'})\rangle$.

Finally, it should be mentioned that in the calculations we can obtain stable solutions when we take $N=n_{max}^a\times n_{max}^b=8\times8=64$ for the system with $\{1234\}\bar{5}$ symmetry, $N=n_{max}^a\times n_{max}^b\times n_{max}^c\times n_{max}^d=6\times6\times6\times6=1296$ for the system with $\{123\}4\bar{5}$ symmetry, while $N=n_{max}^a\times n_{max}^b\times n_{max}^c\times n_{max}^d\times n_{max}^f=5\times5\times5\times5\times5=3125$ for the system with $\{12\}\{34\}\bar{5}$ symmetry.

\begin{table}[htp]
	\setlength{\tabcolsep}{10pt} 
	\begin{center}
		\caption{\label{alpha and rms} The masses and effective harmonic
oscillator parameters $\alpha$ adopted for the final states involving in the fall-apart decays of the all-heavy pentaquarks.}
			\begin{tabular}{clcc}
				\hline\hline
				             State                      & $J^P$ &  Mass (MeV)      & $\alpha_\lambda$/$\alpha_\rho$ (GeV) \\ \hline
				          $\Omega_{ccc}$           & $3/2^+$    &   $4823$    &               $0.5830/0.5049$                \\
				          $\Omega_{bbb}$           & $3/2^+$    &   $14421$    &               $1.1022/0.9545$                \\ \hline
				          $\Omega_{ccb}$           & $1/2^+$    &   $8034$    &               $0.7765/0.5443$                \\
				          $\Omega^*_{ccb}$         & $3/2^+$    &   $8057$    &               $0.7532/0.5342$                \\
				          $\Omega_{cbb}$           & $1/2^+$    &   $11222$    &               $0.7340/0.8911$                \\
				          $\Omega^*_{cbb}$         & $3/2^+$    &   $11250$    &               $0.7074/0.8737$                \\ \hline
                          State                      & $J^P$ &  Mass (MeV)      & $\alpha$ (GeV) \\ \hline				
                          $\eta_c$                 & $0^-$      &  2984~\cite{ParticleDataGroup:2022pth}     &  $0.6654$                              \\
				          $J/\psi$                 & $1^-$      &  3097~\cite{ParticleDataGroup:2022pth}     &  $0.5827$                            \\ \hline
				          $\eta_b$                 & $0^-$      &  9399~\cite{ParticleDataGroup:2022pth}     &  $1.2308$                            \\
				          $\Upsilon(1S)$           & $1^-$      &  9460~\cite{ParticleDataGroup:2022pth}     & $1.1392$                               \\ \hline
				          $B_c$                    & $0^-$      &  6274~\cite{ParticleDataGroup:2022pth}    &  $0.7906$                               \\
				          $B^*_c$                  & $1^-$      &  6328     & $0.7399$                             \\ \hline\hline
			\end{tabular}
	\end{center}
\end{table}

\subsection{Fall-apart decay}

In this work, we calculate the fall-apart decays of the all heavy pentaquarks in a
quark-exchange model~\cite{Barnes:2000hu,Barnes:1991em}. 
In this model, the quark-quark interactions are considered to be
the sources of the fall-apart decays of multiquark states via the
quark rearrangement. The decay amplitude $\mathcal{M}(A\to BC)$ is
described by
\begin{eqnarray}
\mathcal{M}(A\to BC)=-\sqrt{(2\pi)^3}\sqrt{8M_AE_BE_C}\left\langle BC |\sum_{i<j} V_{ij}| A \right\rangle,
\end{eqnarray}
where $A$ stands for the initial multiquark state, $BC$ stands for the final hadron pair.
$V_{ij}$ are the potentials between the inner
quarks of final hadrons $B$ and $C$, they are taken the same as that of
the potential model given in Eq.~(\ref{vij}).
$M_A$ is the mass of the initial state, while $E_B$ and $E_C$ are the energies of the final states $B$ and $C$ in the initial-hadron-rest
system, respectively.
Then, the partial decay width of the $A\to BC$ process is given by
\begin{eqnarray}
\Gamma=\frac{1}{2J_A+1}\frac{|\mathbf{q}|}{8\pi M_A^2}\left|\mathcal{M}(A\to BC)\right|^2,
\end{eqnarray}
where $\mathbf{q}$ is the three momentum of the final state $B$ or $C$ in the initial-hadron-rest
system.

This phenomenological model has been achieved a good description of the low-energy $S$-wave phase shift for the $I=2$ $\pi\pi$
scattering at the quark level~\cite{Barnes:2000hu,Barnes:1991em}. Recently, this model
has also been extended to study the fall-apart decays of the multiquark states in the literature~\cite{Wang:2019spc,Xiao:2019spy,Wang:2020prk,Han:2022fup,Liu:2024fnh,Liu:2022hbk,liu:2020eha}, and a lot of
inspiring results are obtained.
In the present work, the masses and wave functions of the initial pentaquark states are adopted
the numerical results obtained from our potential model calculations.
While for the final $B$ and $C$ hadron states, for simplicity, the wave functions are adopted a single
harmonic oscillator form by fitting their root mean square radii determined within the potential model.
For the unestablished hadrons in the final states, the masses are adopted our quark model predictions,
while for the well established hadrons, the masses are adopted the PDG averaged values.
The harmonic oscillator parameters together with the masses for the final meson and baryon
states are collected in Table~\ref{alpha and rms}.

\begin{figure}
\centering \epsfxsize=8.4 cm \epsfbox{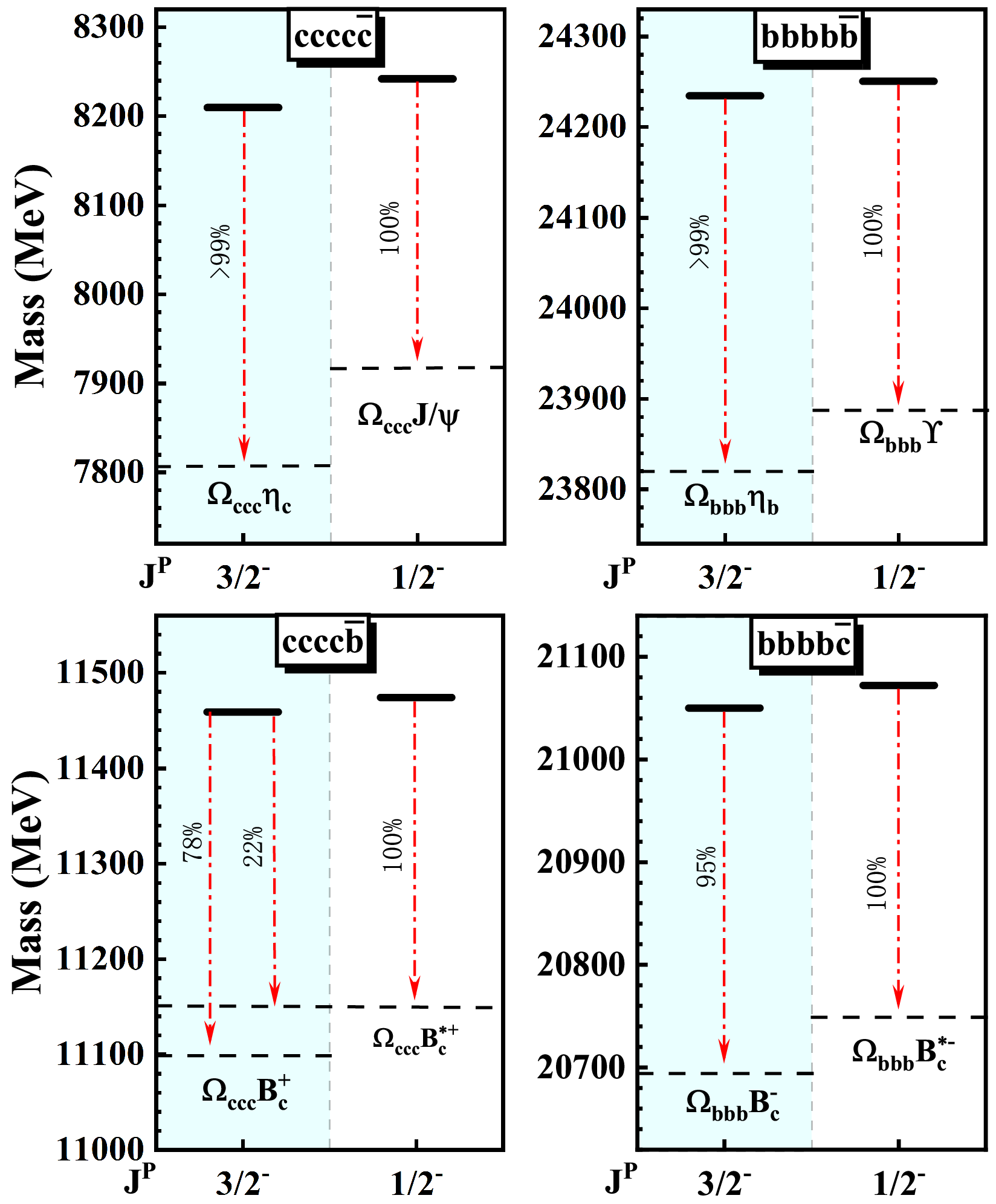} \vspace{-0.1 cm}
\caption{Spectrum for all-heavy pentaquarks with $\{1234\}\bar{5}$ symmetry.
The main decay fall-apart decay channels (dashed lines) and their branching fractions in the total fall-apart decay width for each state are also labeled. }\label{aaa}
\end{figure}

\begin{figure}
\centering \epsfxsize=8.4 cm \epsfbox{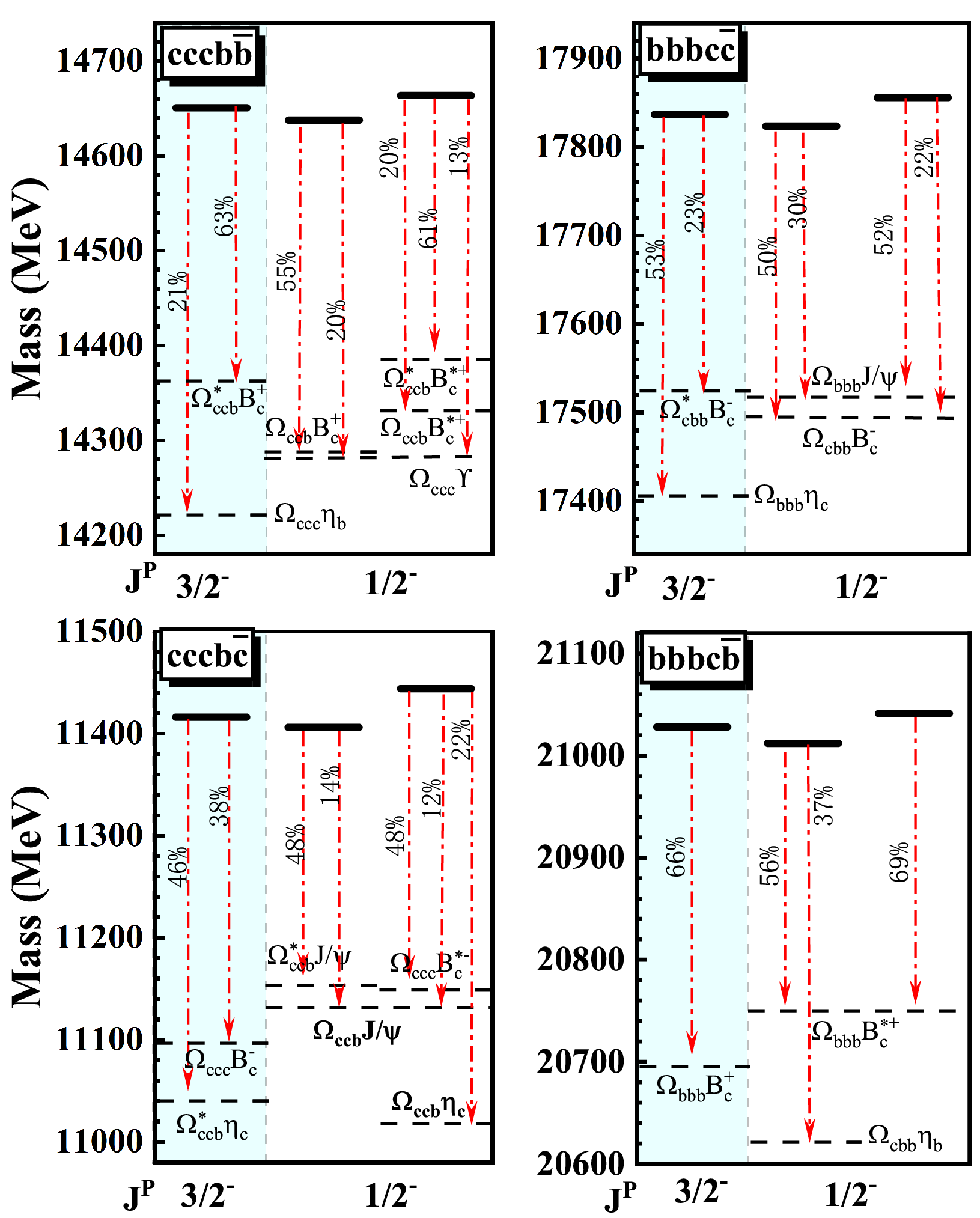} \vspace{-0.1 cm}
\caption{Spectrum (solid lines) for all-heavy pentaquarks with $\{123\}4\bar{5}$ symmetry. The main decay fall-apart decay channels (dashed lines) and their branching fractions in the total fall-apart decay width for each state are also labeled.}\label{aab}
\end{figure}

\begin{figure*}
\centering \epsfxsize=15.8 cm \epsfbox{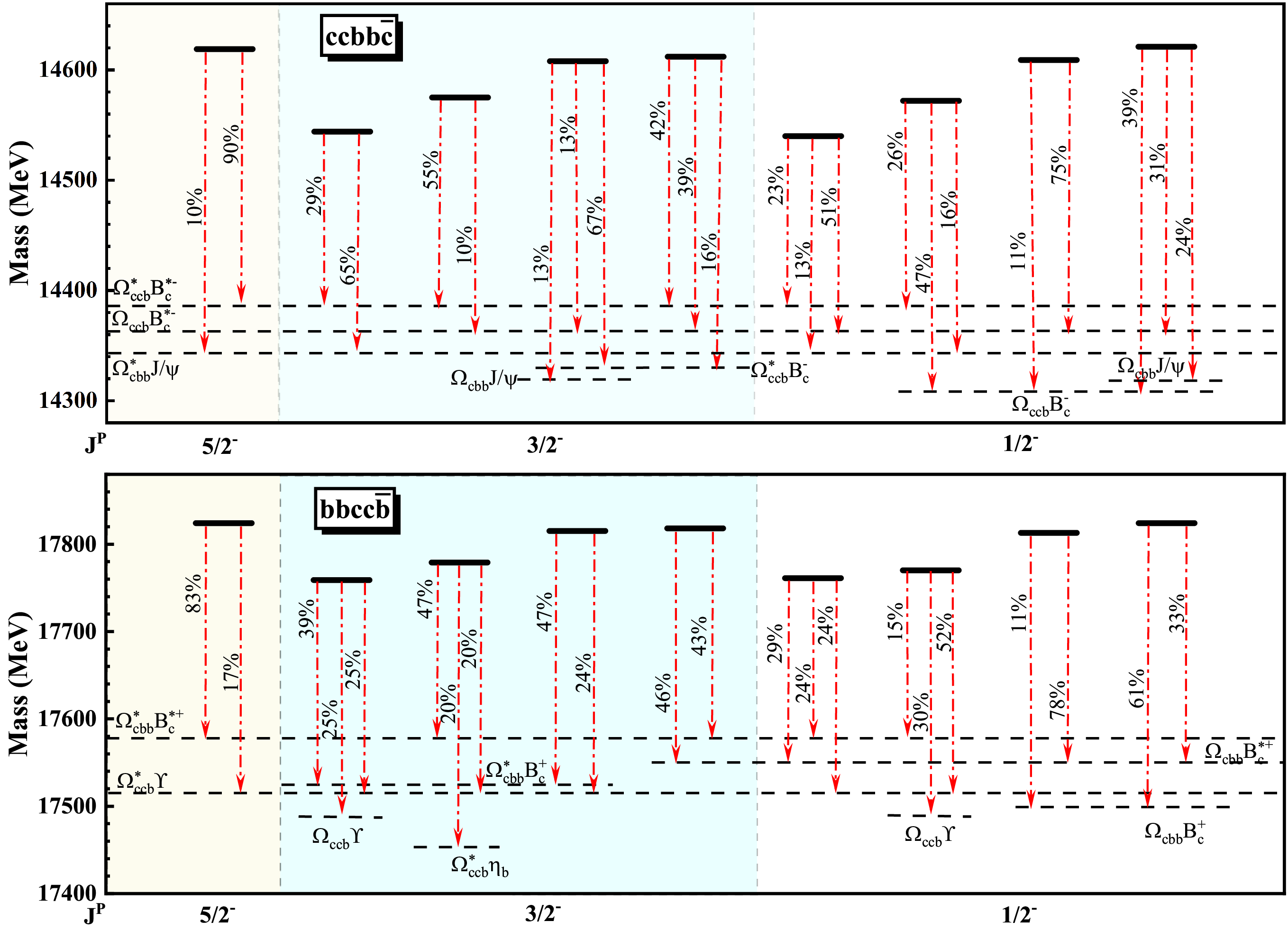} \vspace{-0.1 cm}
\caption{Spectrum (solid lines) for all-heavy pentaquarks with $\{12\}\{34\}\bar{5}$ symmetry. The main decay fall-apart decay channels (dashed lines) and their branching fractions in the total fall-apart decay width for each state are also labeled. }\label{aac}
\end{figure*}

\begin{table*}[hptb]
	\caption{Predicted mass spectra of $1S$ states for the $cccc\bar{c}$, $bbbb\bar{b}$, $cccc\bar{b}$, and $bbbb\bar{c}$ systems with the $\{1234\}\bar{5}$ symmetry compared with those of other works. The unit is MeV.}\label{Comparing the results}
	\tabcolsep=0.15cm
	\renewcommand\arraystretch{1.00}
	\begin{tabular}{cccccccccccc}
		\hline\hline
		& $J^P$ &  Ours &  Ref.~\cite{An:2022fvs} & Ref.~\cite{An:2020jix} & Ref.~\cite{Zhang:2023hmg}&\cite{Gordillo:2024blx}& Ref.~\cite{Rashmi:2024ako} & Ref.~\cite{Yang:2022bfu}& Ref.~\cite{Yan:2021glh} & Ref.~\cite{Wang:2021xao} & Ref.~\cite{Zhang:2020vpz}  \\ \hline
		\multirow{2}*{$cccc\bar{c}$} & $3/2^-$  & 8210    & 8145  & 7864  & 8229  &8151 & 8547   &8095  &             &                & 7410$^{+270}_{-310}$ \\
		                             & $1/2^-$  & 8242    & 8193  & 7949  & 8262  &8194 & 8537   &8045  &  7892  & $7930\pm 150$  &                  \\ \hline
		\multirow{2}*{$bbbb\bar{b}$} & $3/2^-$  & 24235   & 24211 & 23775 & 24761 &24192& 25215  &24035 & 23748 &                & 21600$^{+730}_{-220}$ \\
		                             & $1/2^-$  & 24251   &  24248& 23821 & 24770 &24211& 25213  & 24035 & 23810 & 23910$\pm 150$ &                       \\ \hline
		\multirow{2}*{$cccc\bar{b}$} & $3/2^-$  & 11459   & 11478 & 11130 & 11569 &11417& 11888  &    &             &                &                       \\
		                             & $1/2^-$  & 11474   & 11502 & 11177 & 11582 &11437& 11868  &     &             &                &                       \\ \hline
		\multirow{2}*{$bbbb\bar{c}$} & $3/2^-$  & 21050   & 20975 & 20652 & 21472 &21018& 21879  &     &             &                &                       \\
		                             & $1/2^-$  & 21072   & 21026 & 20699 & 21491 &21046& 21880  &     &             &                &                       \\ \hline\hline
	\end{tabular}
\end{table*}
	
\begin{table*}[hptb]
	\caption{Predicted mass spectra of $1S$ states for the $cccb\bar{c}$, $cccb\bar{b}$, $bbbc\bar{b}$ and $bbbc\bar{c}$ systems with the $\{123\}4\bar{5}$ symmetry, and the $ccbb\bar{c}$ and $bbcc\bar{b}$ systems with the $\{12\}\{34\}\bar{5}$ symmetry.}\label{[123]45 Mixing}
	\begin{tabular}{cccccc}
		\hline\hline	 &\multicolumn{2}{c}{\underline{~~~~~~~~~~~~~~~~~~~~~~~~~~~~$cccb\bar{c}$~~~~~~~~~~~~~~~~~~~~~~~~~}}
		\quad\quad&\multicolumn{2}{c}{\underline{~~~~~~~~~~~~~~~~~~~~~~~~~$cccb\bar{b}$~~~~~~~~~~~~~~~~~~~~~~~~~}}&\\
		~~~~~~$J^P$~~~~~~ & Mass  & ~~~~~~~~~~~~~~~~Eigenvector~~~~~~~~~~~~~~~~ \quad\quad & Mass & ~~~~~~~~~~~~~~~~Eigenvector~~~~~~~~~~~~~~~~ & Configuration\\ \hline	
		$3/2^-$ & 11416 &1 \quad\quad & 14651 &1 &$\begin{pmatrix}
			1S_{\frac{3}{2}^-}(\{123\}4\bar{5})_3
		\end{pmatrix}$ \\ \hline
		$1/2^-$ &
		$\begin{pmatrix}  11406  \\  11444   \end{pmatrix}$ &
		$\begin{pmatrix}
			0.34  & -0.94 \\
			-0.94 & -0.34
		\end{pmatrix}$\quad\quad  &
		$\begin{pmatrix}  14638  \\  14664  \end{pmatrix}$ &
		$\begin{pmatrix}
			0.24  & -0.97   \\
			-0.97 & -0.24
		\end{pmatrix}$&$\begin{pmatrix}
			1S_{\frac{1}{2}^-}(\{123\}4\bar{5})_2\\1S_{\frac{1}{2}^-}(\{123\}4\bar{5})_3
		\end{pmatrix}$  \\ \hline\hline	
		&\multicolumn{2}{c}{\underline{~~~~~~~~~~~~~~~~~~~~~~~~~~~~$bbbc\bar{b}$~~~~~~~~~~~~~~~~~~~~~~~~~}}
		\quad\quad&\multicolumn{2}{c}{\underline{~~~~~~~~~~~~~~~~~~~~~~~~~$bbbc\bar{c}$~~~~~~~~~~~~~~~~~~~~~~~~~}}&\\	
		~~~~~~$J^P$~~~~~~ & Mass  & ~~~~~~~~~~~~~~~~Eigenvector~~~~~~~~~~~~~~~~ \quad\quad & Mass & ~~~~~~~~~~~~~~~~Eigenvector~~~~~~~~~~~~~~~~ & Configuration\\ \hline
		$3/2^-$ &21028 &1  \quad\quad& 17837 &1&$\begin{pmatrix}
			1S_{\frac{3}{2}^-}(\{123\}4\bar{5})_3
		\end{pmatrix}$\\ \hline
		$1/2^-$ &
		$\begin{pmatrix}  21012  \\  21041  \end{pmatrix}$ &
		$\begin{pmatrix}
			0.16  & -0.99  \\
			-0.99 & -0.16
		\end{pmatrix}$  \quad\quad&
		$\begin{pmatrix}  17824  \\  17856  \end{pmatrix}$ &
		$\begin{pmatrix}
			0.16  & -0.99  \\
			-0.99 & -0.16
		\end{pmatrix}$&$\begin{pmatrix}
			1S_{\frac{1}{2}^-}(\{123\}4\bar{5})_2\\1S_{\frac{1}{2}^-}(\{123\}4\bar{5})_3
		\end{pmatrix}$  \\
		\hline\hline
		 &\multicolumn{2}{c}{\underline{~~~~~~~~~~~~~~~~~~~~~~~~~~~~$ccbb\bar{c}$~~~~~~~~~~~~~~~~~~~~~~~~~~~~}}
\quad\quad&\multicolumn{2}{c}{\underline{~~~~~~~~~~~~~~~~~~~~~~~~~~~~$bbcc\bar{b}$~~~~~~~~~~~~~~~~~~~~~~~~~~~~}}&\\
		~~~~~~$J^P$~~~~~~ & Mass   & Eigenvector \quad\quad & Mass  & Eigenvector& Configuration\\ \hline
		$5/2^-$ &14619  &  1   \quad\quad & 17824  &1 &$(1S_{\frac{5}{2}^-}(\{12\}\{34\}\bar{5}))$\\ \hline
		$3/2^-$ & $\begin{pmatrix} 14544  \\  14575  \\  14608  \\14612  \end{pmatrix}$ &
		$\begin{pmatrix}
			-0.46 & -0.07 & 0.83 & -0.30  \\
			0.10  & 0.05  & 0.39 & 0.91  \\
			-0.63  & -0.67  & -0.32 & 0.24 \\
			-0.62 & 0.74  & -0.24  & 0.13
		\end{pmatrix}$  \quad\quad & $\begin{pmatrix}   17759  \\  17779  \\  17815  \\17818 \end{pmatrix}$ &
		$\begin{pmatrix}
			0.25 & -0.05 & -0.90 & 0.35  \\
			0.20  & 0.14  & 0.40 & 0.89  \\
			-0.81  & 0.53  & -0.18 & 0.18 \\
			-0.49 & -0.84  & 0  & 0.24
		\end{pmatrix}$ &$\begin{pmatrix}
			1S_{\frac{3}{2}^-}(\{12\}\{34\}\bar{5})_1\\1S_{\frac{3}{2}^-}(\{12\}\{34\}\bar{5})_2\\
			1S_{\frac{3}{2}^-}(\{12\}\{34\}\bar{5})_3\\1S_{\frac{3}{2}^-}(\{12\}\{34\}\bar{5})_4
		\end{pmatrix}$ \\\hline
		$1/2^-$ &
		$\begin{pmatrix}  14540 \\  14572  \\  14609\\14621  \end{pmatrix}$ &
		$\begin{pmatrix}
			-0.50  & -0.83  & -0.09 & -0.23 \\
			0.35 & -0.23 & 0.87 & -0.27 \\
			0.03 & -0.29 & 0.20 & 0.93  \\
			0.80 & -0.41  & -0.44  & -0.05
		\end{pmatrix}$\quad\quad&
		$\begin{pmatrix}  17761 \\  17770  \\  17813\\17824   \end{pmatrix}$ &
		$\begin{pmatrix}
			-0.38  & -0.91  & -0.12 & -0.09 \\
			-0.35 & 0.24 & -0.88 & 0.20 \\
			-0.08 & 0.15 & -0.15 & -0.97  \\
			0.85 & -0.30  & -0.43  & -0.04
		\end{pmatrix}$&$\begin{pmatrix} 1S_{\frac{1}{2}^-}(\{12\}\{34\}\bar{5})_1\\1S_{\frac{1}{2}^-}(\{12\}\{34\}\bar{5})_2\\
			1S_{\frac{1}{2}^-}(\{12\}\{34\}\bar{5})_3\\1S_{\frac{1}{2}^-}(\{12\}\{34\}\bar{5})_4
		\end{pmatrix}$  \\\hline\hline
	\end{tabular}
\end{table*}

\begin{table*}[hptb]
\caption{The average contributions of each part of the Hamiltonian (in MeV) and the root mean square
radii (in fm) for the $1S$-wave configurations of the all-heavy pentaquarks.}\label{aa}
	\tabcolsep=0.175cm
	\renewcommand\arraystretch{1.00}
	\begin{tabular}{cccccccccccccc}
		\hline\hline
		&\multirow{2}*{$J^P$}& \multirow{2}*{Configuration} & \multirow{2}*{Mass} & \multirow{2}*{$\left\langle T\right\rangle$} & \multirow{2}*{$\left\langle V^{Conf}\right\rangle$} & \multirow{2}*{$\left\langle V^{Coul}\right\rangle$} & \multirow{2}*{$\left\langle V^{SS}\right\rangle$} & $r_{12},r_{13},r_{14},$ & $r_{15},r_{25},$ &&\\
		&&&&&&&& $r_{23},r_{24},r_{34}$  & $r_{35},r_{45}$&&\\ \hline\hline
		\multirow{2}*{$cccc\bar{c}$} & $3/2^-$ & $1S_{\frac{3}{2}^-}(\{1234\}\bar{5})$     & 8210 & 886 & 860.7 & -977.8 & 26.1     & 0.5248 & 0.5176   \\
		                             & $1/2^-$ & $1S_{\frac{1}{2}^-}(\{1234\}\bar{5})$     & 8242 & 846 & 879.8 & -954.9 & 56.3     & 0.5327 & 0.5337   \\ \hline
		\multirow{2}*{$cccc\bar{b}$} & $3/2^-$ & $1S_{\frac{3}{2}^-}(\{1234\}\bar{5})$     & 11459  & 858 &  800  & -1017  &  34      & 0.5135 & 0.4413   \\
		                             & $1/2^-$ & $1S_{\frac{1}{2}^-}(\{1234\}\bar{5})$     & 11474  & 842 &  807  & -1007  &  48      & 0.5172 & 0.4467   \\ \hline
		\multirow{2}*{$bbbb\bar{c}$} & $3/2^-$ & $1S_{\frac{3}{2}^-}(\{1234\}\bar{5})$     & 21050  & 919 &  548  & -1317  &  9       & 0.2988 & 0.3864   \\
		                             & $1/2^-$ & $1S_{\frac{1}{2}^-}(\{1234\}\bar{5})$     & 21072  & 889 &  557  & -1296  &  30      & 0.3017 & 0.3957   \\ \hline
		\multirow{2}*{$bbbb\bar{b}$} & $3/2^-$ & $1S_{\frac{3}{2}^-}(\{1234\}\bar{5})$     & 24235  & 931 &  467  & -1436  &  14      & 0.2842 & 0.2839   \\
		                             & $1/2^-$ & $1S_{\frac{1}{2}^-}(\{1234\}\bar{5})$     & 24251  & 905 &  473  & -1416  &  29      & 0.2870 & 0.2890   \\ \hline\hline
		&$J^P$& Configuration & Mass & $\left\langle T\right\rangle$ & $\left\langle V^{Conf}\right\rangle$ & $\left\langle V^{Coul}\right\rangle$ & $\left\langle V^{SS}\right\rangle$ & $r_{12},r_{13},r_{23}$ & $r_{14},r_{24},r_{34}$ & $r_{15},r_{25},r_{35}$ & $r_{45}$  \\ \hline\hline
		 \multirow{3}*{$cccb\bar{c}$} &        $3/2^-$         & $1S_{\frac{3}{2}^-}(\{123\}4\bar{5})_3$   & 11416 & 897 & 776 & -1058 & 16 &   0.5129 & 0.4302 & 0.5002 & 0.4756     \\
		                                          & \multirow{2}*{$1/2^-$} & $1S_{\frac{1}{2}^-}(\{123\}4\bar{5})_2$   & 11440 & 866 & 790 & -1039 & 39 &   0.5198 & 0.4342 & 0.5125 & 0.4869     \\
		                                          &                        & $1S_{\frac{1}{2}^-}(\{123\}4\bar{5})_3$   & 11411 & 904 & 774 & -1063 & 11 &   0.5105 & 0.4258 & 0.5019 & 0.4753     \\ \hline
		 \multirow{3}*{$cccb\bar{b}$} &        $3/2^-$         & $1S_{\frac{3}{2}^-}(\{123\}4\bar{5})_3$   & 14651 & 887 & 709 & -1121 & 23 &   0.4941 & 0.4185 & 0.4168 & 0.3809     \\
		                                          & \multirow{2}*{$1/2^-$} & $1S_{\frac{1}{2}^-}(\{123\}4\bar{5})_2$   & 14662 & 871 & 715 & -1111 & 34 &   0.4977 & 0.4208 & 0.4216 & 0.3852     \\
		                                          &                        & $1S_{\frac{1}{2}^-}(\{123\}4\bar{5})_3$   & 14640 & 901 & 704 & -1131 & 13 &   0.4906 & 0.4133 & 0.4157 & 0.3777     \\ \hline
		 \multirow{3}*{$bbbc\bar{c}$} &        $3/2^-$         & $1S_{\frac{3}{2}^-}(\{123\}4\bar{5})_3$   & 17837 & 921 & 622 & -1233 & 5  &   0.3309 & 0.3955 & 0.3924 & 0.4838     \\
		                                          & \multirow{2}*{$1/2^-$} & $1S_{\frac{1}{2}^-}(\{123\}4\bar{5})_2$   & 17855 & 896 & 630 & -1215 & 22 &   0.3343 & 0.3984 & 0.4001 & 0.4920     \\
		                                          &                        & $1S_{\frac{1}{2}^-}(\{123\}4\bar{5})_3$   & 17825 & 938 & 617 & -1244 & -7 &   0.3282 & 0.3884 & 0.3933 & 0.4803     \\ \hline
		 \multirow{3}*{$bbbc\bar{b}$} &        $3/2^-$         & $1S_{\frac{3}{2}^-}(\{123\}4\bar{5})_3$   & 21028 & 931 & 541 & -1343 & 9  &   0.3080 & 0.3871 & 0.2894 & 0.4047     \\
		                                          & \multirow{2}*{$1/2^-$} & $1S_{\frac{1}{2}^-}(\{123\}4\bar{5})_2$   & 21040 & 910 & 546 & -1328 & 21 &   0.3110 & 0.3893 & 0.2935 & 0.4086     \\
		                                          &                        & $1S_{\frac{1}{2}^-}(\{123\}4\bar{5})_3$   & 21013 & 951 & 535 & -1357 & -7 &   0.3058 & 0.3795 & 0.2890 & 0.3986     \\ \hline\hline
		&\multirow{2}*{$J^P$}& \multirow{2}*{Configuration} & \multirow{2}*{Mass} & \multirow{2}*{$\left\langle T\right\rangle$} & \multirow{2}*{$\left\langle V^{Conf}\right\rangle$} & \multirow{2}*{$\left\langle V^{Coul}\right\rangle$} & \multirow{2}*{$\left\langle V^{SS}\right\rangle$} & \multirow{2}*{$r_{12}$/$r_{34}$} & $r_{13},r_{14},$ & \multirow{2}*{$r_{15},r_{25}$} &  \multirow{2}*{$r_{35},r_{45}$} \\
		                                         &                        &                                               &       &     &     &       &      &        & $r_{23},r_{24}$ &        &                \\ \hline\hline
		\multirow{9}*{$ccbb\bar{c}$} &        $5/2^-$         & $1S_{\frac{5}{2}^-}(\{12\}\{34\}\bar{5})$     & 14619 & 903 & 703 & -1161 & 23   & 0.4777/0.2898 &     0.4492      & 0.5042 & 0.4225   \\
		                                         & \multirow{4}*{$3/2^-$} & $1S_{\frac{3}{2}^-}(\{12\}\{34\}\bar{5})_1$   & 14595 & 935 & 690 & -1182 & -1   & 0.4722/0.2877 &     0.4431      & 0.4918  & 0.4129   \\
		                                         &                        & $1S_{\frac{3}{2}^-}(\{12\}\{34\}\bar{5})_2$   & 14610 & 915 & 698 & -1170 & 13   & 0.4749/0.2886 &     0.4453      & 0.5001  & 0.4197   \\
		                                         &                        & $1S_{\frac{3}{2}^-}(\{12\}\{34\}\bar{5})_3$   & 14559 & 921 & 676 & -1201 & 10   & 0.5291/0.2876 &     0.4294      & 0.4836  & 0.4849   \\
		                           &                        & $1S_{\frac{3}{2}^-}(\{12\}\{34\}\bar{5})_4$   & 14575 & 924 & 682 & -1191 & 7    & 0.4626/0.3829 &     0.4354      & 0.5164  & 0.3929   \\
		                                         & \multirow{4}*{$1/2^-$} & $1S_{\frac{1}{2}^-}(\{12\}\{34\}\bar{5})_1$   & 14595 & 935 & 690 & -1182 & -1   & 0.4716/0.2874 &     0.4416      & 0.4927  & 0.4139   \\
		                                         &                        & $1S_{\frac{1}{2}^-}(\{12\}\{34\}\bar{5})_2$   & 14561 & 917 & 677 & -1198 & 12   & 0.5300/0.2879 &     0.4305      & 0.4850  & 0.4874   \\
		                                         &                        & $1S_{\frac{1}{2}^-}(\{12\}\{34\}\bar{5})_3$   & 14583 & 912 & 686 & -1183 & 15   & 0.4647/0.3848 &     0.4384      & 0.5220  & 0.3957   \\
		                                         &                        & $1S_{\frac{1}{2}^-}(\{12\}\{34\}\bar{5})_4$   & 14603 & 925 & 694 & -1176 & 6    & 0.4730/0.2879 &     0.4427      & 0.4969  & 0.4173   \\ \hline
		\multirow{9}*{$bbcc\bar{b}$} &        $5/2^-$         & $1S_{\frac{5}{2}^-}(\{12\}\{34\}\bar{5})$     & 17824 & 915 & 623 & -1256 & 20   & 0.2783/0.4642 &     0.4291      & 0.3109  & 0.4202   \\
		                                         & \multirow{4}*{$3/2^-$} & $1S_{\frac{3}{2}^-}(\{12\}\{34\}\bar{5})_1$   & 17811 & 936 & 616 & -1270 & 7    & 0.2764/0.4612 &     0.4254      & 0.3063  & 0.4151   \\
		                                         &                        & $1S_{\frac{3}{2}^-}(\{12\}\{34\}\bar{5})_2$   & 17816 & 928 & 619 & -1264 & 12   & 0.2772/0.4618 &     0.4256      & 0.3091  & 0.4173   \\
		                                         &                        & $1S_{\frac{3}{2}^-}(\{12\}\{34\}\bar{5})_3$   & 17764 & 942 & 601 & -1310 & 9    & 0.3378/0.4612 &     0.4228      & 0.2840  & 0.4379   \\
		                             &                        & $1S_{\frac{3}{2}^-}(\{12\}\{34\}\bar{5})_4$   & 17780 & 934 & 605 & -1294 & 12   & 0.2864/0.4930 &     0.4096      & 0.3834  & 0.3924   \\
		                                         & \multirow{4}*{$1/2^-$} & $1S_{\frac{1}{2}^-}(\{12\}\{34\}\bar{5})_1$   & 17808 & 940 & 615 & -1273 & 4    & 0.2761/0.4599 &     0.4234      & 0.3063  & 0.4142   \\
		                                         &                        & $1S_{\frac{1}{2}^-}(\{12\}\{34\}\bar{5})_2$   & 17768 & 936 & 603 & -1306 & 13   & 0.3385/0.4623 &     0.4244      & 0.2848  & 0.4402   \\
		                                         &                        & $1S_{\frac{1}{2}^-}(\{12\}\{34\}\bar{5})_3$   & 17781 & 932 & 606 & -1292 & 13   & 0.2866/0.4936 &     0.4103      & 0.3849  & 0.3931   \\
		                                         &                        & $1S_{\frac{1}{2}^-}(\{12\}\{34\}\bar{5})_4$   & 17811 & 936 & 616 & -1270 & 7    & 0.2764/0.4602 &     0.4236      & 0.3077  & 0.4153   \\ \hline\hline
	\end{tabular}
\end{table*}

\begin{table*}[htp]
	\setlength{\tabcolsep}{10pt} 
	\begin{center}
		\caption{The predicted partial decay widths of the fall-apart decay processes for the $1S$ states of the $cccc\bar{c}$, $cccc\bar{b}$, $bbbb\bar{b}$ and $bbbb\bar{c}$ systems with the $\{1234\}\bar{5}$ symmetry, and the $cccb\bar{c}$, $cccb\bar{b}$, $bbbc\bar{c}$ and $bbbc\bar{b}$ systems with the $\{123\}4\bar{5}$ symmetry. The unit is MeV.}\label{decaya}
		\begin{tabular}{clcccccc}
			\hline\hline
\multicolumn{4}{c}{$\underline{~~~~~~~~~~~~~~~~~~~~~~~~~~~~~~~~~~~~cccc\bar{c}~~~~~~~~~~~~~~~~~~~~~~~~~~~~~~~~~~~~~~~~~~}$ }&\multicolumn{4}{c}{$\underline{~~~~~~~~~~~~~~~~~~~~~~~~~~~~~~~~~~~~cccc\bar{b}~~~~~~~~~~~~~~~~~~~~~~~~~~~~~~~~~~~~~~~~~~}$}\\
State & $\Omega_{ccc}\eta_c$ &  $\Omega_{ccc}J/\psi$  &       Sum & State  & $\Omega_{ccc}B^+_c$  & $\Omega_{ccc}B^{*+}_c$ &Sum\\ \hline
$P_{c^4\bar{c}}(8210)3/2^-$  &        $2.4$         &     $8.9\times10^{-4}$     &2.40  & $P_{c^4\bar{b}}(11459)3/2^-$ &        $1.27$        &         $0.35$  &  1.62     \\
$P_{c^4\bar{c}}(8242)1/2^-$  &       $\cdots$       &         $3.76$         & 3.76 & $P_{c^4\bar{b}}(11474)1/2^-$ &       $\cdots$       &         $2.57$  & 2.57      \\ \hline\hline
\multicolumn{4}{c}{$\underline{~~~~~~~~~~~~~~~~~~~~~~~~~~~~~~~~~~~~bbbb\bar{c}~~~~~~~~~~~~~~~~~~~~~~~~~~~~~~~~~~~~~~~~~~}$ }&\multicolumn{4}{c}{$\underline{~~~~~~~~~~~~~~~~~~~~~~~~~~~~~~~~~~~~bbbb\bar{b}~~~~~~~~~~~~~~~~~~~~~~~~~~~~~~~~~~~~~~~~~~}$}\\
State                 & $\Omega_{bbb}B^-_c$  & $\Omega_{bbb}B^{*-}_c$ &  Sum   & State   & $\Omega_{bbb}\eta_b$ & $\Omega_{bbb}\Upsilon$ &Sum\\ \hline
$P_{b^4\bar{c}}(21050)3/2^-$ &        $0.69$        &         $0.03$         & 0.72 & $P_{b^4\bar{b}}(24235)3/2^-$ &        $0.55$        &     $4.8\times10^{-6}$ & 0.55   \\
$P_{b^4\bar{c}}(21072)1/2^-$ &       $\cdots$       &         $0.96$         & 0.96  & $P_{b^4\bar{b}}(24251)1/2^-$ &       $\cdots$       &         $0.84$  & 0.84      \\ \hline\hline
\multicolumn{8}{c}{$cccb\bar{c}$ }\\ \hline
			                              State                   & $\Omega_{ccc}B^-_c$  & $\Omega_{ccc}B^{*-}_c$ & $\Omega_{ccb}\eta_c$ &  $\Omega_{ccb}J/\psi$  & $\Omega^*_{ccb}\eta_c$ &  $\Omega^*_{ccb}J/\psi$ & Sum \\ \hline
			  $P_{c^3b\bar{c}}(11416)3/2^-$   &        $0.56$        &         $0.17$         &       $\cdots$       &     $5.19\times10^{-4}$     &         $0.69$         &          $0.05$     &  1.47   \\
			                              $P_{c^3b\bar{c}}(11406)1/2^-$ &       $\cdots$       &         $0.22$         &        $0.29$        &         $0.20$         &        $\cdots$        &          $0.67$   &   1.38    \\
			                              $P_{c^3b\bar{c}}(11444)1/2^-$ &       $\cdots$       &         $1.02$         &        $0.48$        &         $0.27$         &        $\cdots$        &          $0.35$   &  2.12     \\ \hline\hline
\multicolumn{8}{c}{$cccb\bar{b}$ }\\ \hline
			                              State                   & $\Omega_{ccc}\eta_b$ & $\Omega_{ccc}\Upsilon$ & $\Omega_{ccb}B^+_c$  & $\Omega_{ccb}B^{*+}_c$ & $\Omega^*_{ccb}B^+_c$  & $\Omega^*_{ccb}B^{*+}_c$ &Sum\\ \hline
			  $P_{c^3b\bar{b}}(14651)3/2^-$   &        $0.15$        &     $8.31\times10^{-7}$     &       $\cdots$       &         $0.04$         &         $0.45$         &          $0.07$     &  0.71   \\
			                              $P_{c^3b\bar{b}}(14638)1/2^-$ &       $\cdots$       &         $0.18$         &        $0.49$        &         $0.22$         &        $\cdots$        &      $1.1\times10^{-3}$  &   0.89  \\
			                              $P_{c^3b\bar{b}}(14664)1/2^-$ &       $\cdots$       &         $0.18$         &        $0.06$        &         $0.26$         &        $\cdots$        &          $0.80$   &  1.30     \\ \hline\hline
\multicolumn{8}{c}{$bbbc\bar{c}$ }\\ \hline
			                              State                   & $\Omega_{bbb}\eta_c$ &  $\Omega_{bbb}J/\psi$  & $\Omega_{cbb}B^-_c$  & $\Omega_{cbb}B^{*-}_c$ & $\Omega^*_{cbb}B^-_c$  & $\Omega^*_{cbb}B^{*-}_c$ &Sum\\ \hline
			  $P_{b^3c\bar{c}}(17837)3/2^-$   &        $0.30$        &     $1.88\times10^{-5}$     &       $\cdots$       &     $1.8\times10^{-3}$      &         $0.13$         &          $0.13$     &   0.56  \\
			                              $P_{b^3c\bar{c}}(17824)1/2^-$ &       $\cdots$       &         $0.09$         &        $0.15$        &     $1.1\times10^{-3}$      &        $\cdots$        &          $0.06$ &    0.30     \\
			                              $P_{b^3c\bar{c}}(17856)1/2^-$ &       $\cdots$       &         $0.52$         &        $0.22$        &         $0.08$         &        $\cdots$        &          $0.17$ &      0.99   \\ \hline\hline
\multicolumn{8}{c}{$bbbc\bar{b}$ }\\ \hline
			                              State                   & $\Omega_{bbb}B^+_c$  & $\Omega_{bbb}B^{*+}_c$ & $\Omega_{cbb}\eta_b$ & $\Omega_{cbb}\Upsilon$ & $\Omega^*_{cbb}\eta_b$ & $\Omega^*_{cbb}\Upsilon$ &Sum\\ \hline
			  $P_{b^3c\bar{b}}(21028)3/2^-$   &        $0.27$        &         $0.01$         &       $\cdots$       &     $4.07\times10^{-4}$     &         $0.06$         &          $0.07$   &   0.41    \\
			                              $P_{b^3c\bar{b}}(21012)1/2^-$ &       $\cdots$       &         $0.18$         &        $0.12$        &     $3.0\times10^{-3}$      &        $\cdots$        &          $0.02$  &   0.32     \\
			                              $P_{b^3c\bar{b}}(21041)1/2^-$ &       $\cdots$       &         $0.36$         &        $0.04$        &         $0.06$         &        $\cdots$        &          $0.06$  &    0.52    \\ \hline\hline
		\end{tabular}																	
	\end{center}
\end{table*}

\begin{table*}[htp]
	\setlength{\tabcolsep}{7pt} 
	\begin{center}
		\caption{The predicted partial decay widths of the fall-apart decay processes for the $1S$ states of the $ccbb\bar{c}$ and $bbcc\bar{b}$ systems with the $\{12\}\{34\}\bar{5}$ symmetry. The unit is MeV.}\label{decayb}
		\begin{tabular}{cccccccccccc}
			\hline\hline
\multicolumn{10}{c}{$ccbb\bar{c}$ }\\ \hline
			                              State                   & $\Omega_{ccb}B^-_c$ & $\Omega_{ccb}B^{*-}_c$ & $\Omega^*_{ccb}B^-_c$ & $\Omega^*_{ccb}B^{*-}_c$ & $\Omega_{cbb}\eta_c$ &  $\Omega_{cbb}J/\psi$  & $\Omega^*_{cbb}\eta_c$ &  $\Omega^*_{cbb}J/\psi$ & Sum \\ \hline
			  $P_{c^2b^2\bar{c}}(14619)5/2^-$   &      $\cdots$       &        $\cdots$        &       $\cdots$        &          $0.95$          &       $\cdots$       &        $\cdots$        &        $\cdots$        &          $0.11$       &   1.06\\
			                              $P_{c^2b^2\bar{c}}(14544)3/2^-$ &      $\cdots$       &         $0.01$         &     $1.5\times10^{-3}$     &          $0.05$          &       $\cdots$       &     $4.4\times10^{-5}$     &      $3.0\times10^{-4}$       &          $0.11$         & 0.17\\
			                              $P_{c^2b^2\bar{c}}(14575)3/2^-$ &      $\cdots$       &         $0.05$         &         $0.1$         &          $0.27$          &       $\cdots$       &         $0.01$         &         $0.04$         &          $0.02$        & 0.49 \\
			                              $P_{c^2b^2\bar{c}}(14608)3/2^-$ &      $\cdots$       &         $0.10$         &        $0.52$         &          $0.01$          &       $\cdots$       &         $0.10$         &         $0.04$         &          $0.01$        & 0.78 \\
			                              $P_{c^2b^2\bar{c}}(14612)3/2^-$ &      $\cdots$       &         $0.67$         &        $0.27$         &          $0.73$          &       $\cdots$       &         $0.02$         &      $3.0\times10^{-4}$       &          $0.03$        &  1.72\\
			                              $P_{c^2b^2\bar{c}}(14540)1/2^-$ &       $0.17$        &         $0.94$         &       $\cdots$        &          $0.43$          &        $0.02$        &         $0.04$         &        $\cdots$        &          $0.24$        &  1.84\\
			                              $P_{c^2b^2\bar{c}}(14572)1/2^-$ &       $0.09$        &         $0.02$         &       $\cdots$        &          $0.05$          &    $1.7\times10^{-3}$     &     $2.2\times10^{-3}$      &        $\cdots$        &          $0.03$        & 0.19 \\
			                              $P_{c^2b^2\bar{c}}(14609)1/2^-$ &       $0.09$        &         $0.59$         &       $\cdots$        &          $0.02$          &        $0.01$        &         $0.03$         &        $\cdots$        &          $0.05$        & 0.79 \\
			                              $P_{c^2b^2\bar{c}}(14621)1/2^-$ &       $0.20$        &         $0.16$         &       $\cdots$        &          $0.01$          &        $0.01$        &         $0.12$         &        $\cdots$        &          $0.01$       &  0.51 \\ \hline\hline
\multicolumn{10}{c}{$bbcc\bar{b}$ }\\ \hline
			                              State                   & $\Omega_{cbb}B^+_c$ & $\Omega_{cbb}B^{*+}_c$ & $\Omega^*_{cbb}B^+_c$ & $\Omega^*_{cbb}B^{*+}_c$ & $\Omega_{ccb}\eta_b$ & $\Omega_{ccb}\Upsilon$ & $\Omega^*_{ccb}\eta_b$ & $\Omega^*_{ccb}\Upsilon$ & Sum\\ \hline
			 $P_{c^2b^2\bar{b}}(17824)5/2^-$   &      $\cdots$       &        $\cdots$        &       $\cdots$        &          $0.68$          &       $\cdots$       &        $\cdots$        &        $\cdots$        &          $0.14$       &   0.82\\
			                              $P_{c^2b^2\bar{b}}(17759)3/2^-$ &      $\cdots$       &     $4.1\times10^{-5}$     &        $0.14$         &          $0.01$          &       $\cdots$       &         $0.09$         &         $0.03$         &          $0.09$      &    0.36\\
			                              $P_{c^2b^2\bar{b}}(17779)3/2^-$ &      $\cdots$       &         $0.01$         &     $2.5\times10^{-3}$     &          $0.07$          &       $\cdots$       &         $0.01$         &         $0.03$         &          $0.03$      &    0.15\\
			                              $P_{c^2b^2\bar{b}}(17815)3/2^-$ &      $\cdots$       &         $0.03$         &        $0.24$         &          $0.04$          &       $\cdots$       &         $0.04$         &         $0.04$         &          $0.12$       &   0.51\\
			                              $P_{c^2b^2\bar{b}}(17818)3/2^-$ &      $\cdots$       &         $0.29$         &        $0.04$         &          $0.27$          &       $\cdots$       &         $0.01$         &         $0.02$         &       $5.0\times10^{-4}$    &    0.63\\
			                              $P_{c^2b^2\bar{b}}(17761)1/2^-$ &       $0.05$        &         $0.14$         &       $\cdots$        &          $0.12$          &        $0.02$        &         $0.04$         &        $\cdots$        &          $0.12$       &  0.49 \\
			                              $P_{c^2b^2\bar{b}}(17770)1/2^-$ &     $1.0\times10^{-4}$     &     $4.0\times10^{-5}$     &       $\cdots$        &          $0.04$          &        $0.01$        &         $0.08$         &        $\cdots$        &          $0.14$       &   0.27\\
			                              $P_{c^2b^2\bar{b}}(17813)1/2^-$ &       $0.04$        &         $0.29$         &       $\cdots$        &          $0.01$          &     $5.0\times10^{-4}$      &         $0.03$         &        $\cdots$        &       $8.0\times10^{-4}$    & 0.37   \\
			                              $P_{c^2b^2\bar{b}}(17824)1/2^-$ &       $0.20$        &         $0.11$         &       $\cdots$        &          $0.01$          &    $4.1\times10^{-3}$     &         $0.02$         &        $\cdots$        &      $1.9\times10^{-3}$    & 0.33  \\ \hline\hline
		\end{tabular}																	
	\end{center}
\end{table*}

\section{RESULTS AND DISCUSSIONS}\label{RESULTS AND DISCUSSIONS}

\subsection{Pentaquarks with $\{1234\}\bar{5}$ symmetry}\label{results of Scheme 1}

In the all-heavy pentaquarks, the $cccc\bar{c}$, $cccc\bar{b}$, $bbbb\bar{c}$, and $bbbb\bar{b}$
systems have the $\{1234\}\bar{5}$ symmetry. Considering the permutation symmetry
of identical quarks, there are two states with $J^P=3/2^-$ and $J^P=1/2^-$ for each pentaquark system.

\subsubsection{Mass}

The predicted mass spectrum has been given in Table~\ref{Comparing the results}
and also shown in Fig.~\ref{aaa}. For a comparison, some predictions of other works are also collected in Table~\ref{Comparing the results}.

For the $cccc\bar{c}$ system, the masses of the ground ($1S$) states are predicted to be $\sim8.2$ GeV, which is
$\sim300$ MeV above the mass threshold of $J/\psi \Omega_{ccc}(3/2^+)$.
For the $bbbb\bar{b}$ system, the masses of the ground ($1S$) states are predicted to be $\sim24.2$ GeV, which is
$\sim350$ MeV above the mass threshold of $\Omega_{bbb}(3/2^+)\Upsilon$. For the $cccc\bar{b}$ and $bbbb\bar{c}$ systems, the masses of the ground ($1S$) states are predicted to be $\sim11.5$ GeV and $\sim21.1$ GeV, respectively, which are about 300 MeV
above the mass thresholds of $\Omega_{ccc}(3/2^+)B_c^*$ and $\Omega_{bbb}(3/2^+)B_c^*$,
respectively. Our predictions are consistent with those from
the potential model calculations by using the variational method~\cite{An:2022fvs} and the diffusion Monte Carlo method~\cite{Gordillo:2024blx}.
However, our predictions are very different from the predictions from the other works~\cite{Zhang:2023hmg,An:2020jix,Wang:2021xao,Zhang:2020vpz,Yan:2021glh,Rashmi:2024ako}.
For example, our predicted mass for the $cccc\bar{c}$ system much larger than
those predicted within the CMI model~\cite{An:2020jix}, the chiral quark model/quark delocalization
color screening model~\cite{Yan:2021glh}, and the QCD sum rule approaches~\cite{Wang:2021xao,Zhang:2020vpz},
the discrepancy is over 300 MeV.

We also analyze the contributions from each part of
the Hamiltonian for the pentaquark states.
The results are listed in Table~\ref{aa} as well. It shows that the averaged kinetic energy
$\langle T\rangle$ is nearly a stable value $\sim 900$ MeV for all of the states.
The color-Coulomb potential contributes to a large negative value $\langle V^{Coul}\rangle$, about $-950, -1000, -1300, -1400$ MeV
for the $cccc\bar{c}$, $cccc\bar{b}$, $bbbb\bar{c}$, and $bbbb\bar{b}$
systems, respectively. The linear confining potential $\langle V^{Conf}\rangle$ has the same order
of magnitude as $\langle T\rangle$. For the $cccc\bar{c}$ and $cccc\bar{b}$ systems,
the value of confining potential $\langle V^{Conf}\rangle$, $\sim 850$ MeV,
is comparable with that of $\langle T\rangle$. While for the $bbbb\bar{b}$ and $bbbb\bar{c}$ systems,
the confining potential $\langle V^{Conf}\rangle$, $\sim 500$ MeV, is about one
half of the $\langle T\rangle$. It is interesting to find that for a pentaquark system
the value of $\langle T\rangle+\langle V^{Conf}\rangle+\langle V^{Coul}\rangle$
for the $J^P=1/2^-$ state is equal to that of the $J^P=3/2^-$ state,
although each parts for these two states are different.
From this point of view, the mass splitting between the $J^P=1/2^-$
and $3/2^-$ states is caused by the chromo-magnetic interaction.

To know some details of the inner structure of the predicted pentaquark states,
we calculate the root mean square (RMS) radii between two quarks, i.e., $\sqrt{\langle r_{ij}^2\rangle}$.
The results are listed in Tables~\ref{aa}.
It shows that the $1S$ states with $\{1234\}\bar{5}$ symmetry should have a compact structure.
The RMS radii for the $cccc\bar{c}$, $cccc\bar{b}$, $bbbb\bar{c}$,
and $bbbb\bar{b}$ systems are in the ranges of $(0.51,0.54)$, $(0.44,0.52)$, $(0.29,0.40)$,
and $(0.28,0.29)$ fm, respectively.

\subsubsection{Decay}

Since all of the predicted $1S$-wave states are above some dissociation baryon-meson thresholds, we further evaluate the
their fall-apart decay properties. Our results are given in Table~\ref{decaya}. It is found that
the $1S$ pentaquark states with $\{1234\}\bar{5}$ symmetry might be very narrow states,
their fall-apart decay widths scatter in the range of $\sim0.5-4.0$ MeV.

For the $cccc\bar{c}$ system, the $J^P=3/2^-$ state
$P_{c^4\bar{c}}(8210)3/2^-$ may have a fall-apart decay width of a few MeV,
which is nearly saturated by the $\Omega_{ccc}\eta_c$ channel.
The decay rate of the $\Omega_{ccc}J/\psi$ channel is tiny. The partial width ratio between
these two channels is predicted to be
\begin{eqnarray}
	\Gamma[\Omega_{ccc}J/\psi]:\Gamma[\Omega_{ccc}\eta_c]\simeq 2.6\times 10^{-3}.
\end{eqnarray}
While for the $J^P=1/2^-$ state $P_{c^4\bar{c}}(8242)1/2^-$, the $\Omega_{ccc}J/\psi$
is the only fall-apart decay channel, the partial width
is predicted to be
\begin{eqnarray}
	\Gamma[\Omega_{ccc}J/\psi]\simeq 3.76~ \mathrm{MeV}.
\end{eqnarray}
The $\Omega_{ccc}\eta_c$ and $\Omega_{ccc}J/\psi$ may be ideal channels
for searching for the all-charmed pentaquark states with $J^P=3/2^-$ and $J^P=1/2^-$, respectively.

For the $1S$ $bbbb\bar{b}$ states, the fall-apart decay widths are narrower than
the $1S$ $cccc\bar{c}$ states due to the suppression of the heavy bottom quark.
The $P_{b^4\bar{b}}(24235)3/2^-$ dominantly decays into the $\Omega_{bbb}\eta_b$ channel
with a partial width of $\Gamma[\Omega_{bbb}\eta_b]\simeq 0.55$ MeV,
while the decay rate into the $\Omega_{bbb}\Upsilon$ channel is very small.
Our predicted partial width ratio between $\Omega_{bbb}\Upsilon$ and $\Omega_{bbb}\eta_b$,
\begin{eqnarray}
	\Gamma[\Omega_{bbb}\Upsilon]:\Gamma[\Omega_{bbb}\eta_b]\simeq8.7\times 10^{-6},
\end{eqnarray}
is too tiny to be comparable with the value $0.4$ estimated in Ref.~\cite{An:2020jix}.
For the $J^P=1/2^-$ state $P_{b^4\bar{b}}(24251)1/2^-$, the fall-apart decays are saturated
by the $\Omega_{bbb}\Upsilon$ channel, its partial width is predicted to be
\begin{eqnarray}
	\Gamma[\Omega_{bbb}\Upsilon]\simeq 0.84~ \mathrm{MeV}.
\end{eqnarray}
Similarly, the $\Omega_{bbb}\eta_b$ and $\Omega_{bbb}\Upsilon$ may be ideal channels
for searching for the all-bottom pentaquark states with $J^P=3/2^-$ and $J^P=1/2^-$, respectively.

For the $cccc\bar{b}$ system, the $J^P=3/2^-$ state $P_{c^4\bar{b}}(11459)3/2^-$ may have
a fall-apart decay width of a few MeV.
This state can decay into both the $\Omega_{ccc}B_c^{+}$ and $\Omega_{ccc}B_c^{*+}$ channels.
Our predicted partial width ratio between this two channels,
\begin{eqnarray}
	\Gamma[\Omega_{ccc}B_c^{*+}]:\Gamma[\Omega_{ccc}B_c^{+}]\simeq 0.28,
\end{eqnarray}
is comparable with the ratio $0.4$ estimated in Ref.~\cite{An:2020jix}. For the $J^P=1/2^-$
state $P_{c^4\bar{b}}(11474)3/2^-$, the $\Omega_{ccc}B_c^{*+}$ is the only
fall-apart decay channel. The partial decay width is predicted to be
\begin{eqnarray}
    \Gamma[\Omega_{ccc}B_c^{*+}]\simeq 2.57~ \mathrm{MeV}.
\end{eqnarray}
The $\Omega_{ccc}B_c^{+}$ and $\Omega_{ccc}B_c^{*+}$ may be ideal channels for searching
for the $1S$ states with $J^P=3/2^-$ and $J^P=1/2^-$ for the $cccc\bar{b}$ system, respectively.

For the $bbbb\bar{c}$ system, the fall-apart decay widths of the $1S$ states are
about a factor 3 narrower than the $1S$ $cccc\bar{b}$ states due to the suppression of the heavy bottom quark.
The $J^P=3/2^-$ state $P_{b^4\bar{c}}(21050)3/2^-$ dominantly decays into the
$\Omega_{bbb}B_c^{-}$ channel with a partial width of $\Gamma[\Omega_{bbb}B_c^{-}]\simeq 0.69$ MeV.
The decay rate into the $\Omega_{bbb}B_c^{*-}$ channel is tiny.
Our predicted partial width ratio between the $\Omega_{bbb}B_c^{*-}$ and $\Omega_{bbb}B_c^{-}$ channels,
\begin{eqnarray}
    \Gamma[\Omega_{bbb}B_c^{*-}]:\Gamma[\Omega_{bbb}B_c^{-}]\simeq 0.04,
\end{eqnarray}
is about one order of magnitude smaller than that predicted in Ref.~\cite{An:2020jix}.
While for the $J^P=1/2^-$ state $P_{b^4\bar{c}}(21072)1/2^-$,
the $\Omega_{bbb}B_c^{*-}$ is the only fall-apart decay channel.
Its partial width is predicted to be
\begin{eqnarray}
    \Gamma[\Omega_{bbb}B_c^{*-}]\simeq 0.96~ \mathrm{MeV}.
\end{eqnarray}
To look for the $bbbb\bar{c}$ states, the $\Omega_{bbb}B_c^{-}$ and $\Omega_{bbb}B_c^{*-}$
final states are worth observing in future experiments.

\subsection{Pentaquarks with $\{123\}4\bar{5}$ symmetry}

In the all-heavy pentaquarks, the $cccb\bar{b}$, $cccb\bar{c}$, $bbbc\bar{c}$, and $bbbc\bar{b}$
systems have the $\{123\}4\bar{5}$ symmetry. Considering the permutation symmetry
of identical quarks, for each pentaquark system there are seven $1S$ states,
one with $J^P=5/2^-$, three with $J^P=3/2^-$, and three with $J^P=1/2^-$.

\subsubsection{Mass}

Our predicted mass spectrum are listed in Table~\ref{[123]45 Mixing} and also shown in Fig.~\ref{aab}.
It should be pointed out that the $J^P=5/2^-$ configuration $C_1S_1$, two $J^P=3/2^-$ configurations $C_1S_2$ and  $C_1S_3$, and one $J^P=1/2^-$ configuration $C_1S_6$ have a simple color structure $C_1$. In these four special cases, the $Q_1Q_2Q_3$
composed by identical quarks and $Q_4\bar{Q}_5$ are two colorless subsystems. Between $Q_1Q_2Q_3$ and $Q_4\bar{Q}_5$ there are neither short-range interactions arising from one-gluon exchanges nor long-range interactions arising from one-light-meson exchanges.
These states cannot form compact states due to the absence of the OGE potentials.
Thus, they are absent in our predictions. Although there are no one-light-meson exchanged interactions,
they may form loose molecular states due to some other dynamical mechanisms, such as
fully-heavy vector meson exchanges~\cite{Liu:2023gla,Liu:2024pio}, chromopolarizability~\cite{Brambilla:2015rqa,Dong:2022rwr,Dong:2021lkh},
and so on. This may be an interesting topic for future study.

The other three configurations $1S_{\frac{3}{2}^-}(\{123\}4\bar{5})_3$, $1S_{\frac{1}{2}^-}(\{123\}4\bar{5})_2$,
and $1S_{\frac{1}{2}^-}(\{123\}4\bar{5})_3$ may form compact structures due to OGE interactions.
The masses for the $cccb\bar{c}$, $cccb\bar{b}$, $bbbc\bar{c}$, and $bbbc\bar{b}$ systems
are predicted to be $\sim11.4$, $\sim14.6$, $17.8$, and $\sim21.0$ GeV,
respectively. All of the $1S$ states are far above the low-lying dissociation baryon-meson thresholds.
It should be mentioned that there is
a slight mixing between the two $J^P=1/2^-$ configurations (see Table~\ref{[123]45 Mixing}).
About the $cccb\bar{c}$, $cccb\bar{b}$, $bbbc\bar{c}$, and $bbbc\bar{b}$ systems,
some other studies have been carried out within the MIT bag model~\cite{Zhang:2023hmg},
chromo-magnetic interaction model~\cite{An:2020jix}, effective mass and screened charge model~\cite{Rashmi:2024ako},
and various potential models~\cite{Yan:2023kqf,An:2022fvs,Gordillo:2024blx}. There exist large differences in the predictions from various models.
Our predicted mass ranges for the $1S$ states are
comparable with those predicted in Ref.~\cite{An:2022fvs}.

The contributions from each part of the Hamiltonian for the pentaquark states
are listed in Table~\ref{aa} as well.
Taking the $cccb\bar{c}$ system as an example, one can see that the kinetic energy $\langle T \rangle\sim 900$ MeV, the confining
potential $\langle V^{Conf} \rangle\sim 780$ MeV, and the color-Coulomb potential
$\langle V^{Coul}\rangle\sim 1050$ MeV have the same order of magnitude. The sum of different contributions from various states is the same value 615 MeV, thus the mass splitting between different configurations is completely determined by the spin-spin potential.

To know about the inner structure of the pentaquark states, in Table~\ref{aa} we further give the
root mean square (RMS) radii between two quarks. It is found that
the three $1S$ states with $J^P=1/2^-$ and $J^P=3/2^-$ for a given pentaquark system
are compact states. The RMS radii for the $cccb\bar{c}$, $cccb\bar{b}$, $bbbc\bar{c}$,
and $bbbc\bar{b}$ systems are in the ranges of $(0.42,0.52)$, $(0.38,0.50)$, $(0.32,0.50)$,
and $(0.29,0.41)$ fm, respectively.

\subsubsection{Decay}

Considering the fact that all of the predicted $1S$-wave states are above some dissociation baryon-meson thresholds,
by using the masses and wave functions obtained from the potential model we further evaluate the
the fall-apart decay properties of the $1S$ pentaquark states. Our results are given in Table~\ref{decaya}. It is found that
the $1S$ states with $\{123\}4\bar{5}$ symmetry might be narrow states,
their fall-apart decay widths scatter in the range of $\sim0.3-2.0$ MeV.

For the $cccb\bar{c}$ system, the $J^P=3/2^-$ state $P_{c^3b\bar{c}}(11416)3/2^-$ may have
a fall-apart width of $\Gamma\simeq 1.47$ MeV.
This state dominantly decays into the $\Omega_{ccb}^*\eta_c$ and $\Omega_{ccc}B_c^{-}$ and  channels,
and also has sizeable decay rate into the $\Omega_{ccc}B_c^{*-}$ channel.
The partial width ratios between these channels are estimated to be
\begin{eqnarray}
    \Gamma[\Omega_{ccb}^*\eta_c]:\Gamma[\Omega_{ccc}B_c^{-}]:\Gamma[\Omega_{ccc}B_c^{*-}]\simeq 4.0:3.3:1.
\end{eqnarray}
While for the two $J^P=1/2^-$ states, their fall-apart widths are comparable with
that of the $J^P=3/2^-$ state. The low-lying $J^P=1/2^-$ state $P_{c^3b\bar{c}}(11406)1/2^-$ dominantly decays
into the $\Omega_{ccb}^*J/\psi$ channel, and also has sizeable decay rates into $\Omega_{ccb}J/\psi$,
$\Omega_{ccb}\eta_c$ and $\Omega_{ccc}B_c^{*-}$ channels with a comparable partial width.
The partial width ratios between theses channels are estimated to be
\begin{eqnarray}
    \Gamma[\Omega_{ccb}^*J/\psi]:\Gamma[\Omega_{ccb}\eta_c]:\Gamma[\Omega_{ccc}B_c^{*-}]:\Gamma[\Omega_{ccb}J/\psi]\nonumber\\
    \simeq 3.3:1.4:1.1:1.
\end{eqnarray}
The high-lying $J^P=1/2^-$ state $P_{c^3b\bar{c}}(11444)1/2^-$ dominantly decays into the
$\Omega_{ccc}B_c^{*-}$ channel, and also has sizeable decay rates into $\Omega_{ccb}\eta_c$, $\Omega_{ccb}J/\psi$,
and $\Omega_{ccb}^*J/\psi$ channels. The partial width ratios between these main decay channels
are predicted to be
\begin{eqnarray}
    \Gamma[\Omega_{ccc}B_c^{*-}]:\Gamma[\Omega_{ccb}\eta_c]:\Gamma[\Omega_{ccb}^*J/\psi]:\Gamma[\Omega_{ccb}J/\psi]\nonumber\\
    \simeq 3.8:1.8:1.3:1.
\end{eqnarray}
To look for the $cccb\bar{c}$ states, the $\Omega_{ccc}B_c^{(*)-}$,
$\Omega_{ccb}^{(*)}\eta_c$, and $\Omega_{ccb}^{(*)}J/\psi$ channels
are worth observing in experiments.

For the $cccb\bar{b}$ system, the $J^P=3/2^-$ state $P_{c^3b\bar{b}}(14651)3/2^-$ may have a fall-apart width of $\Gamma\simeq 0.71$ MeV.
The fall-apart decays are governed by both the $\Omega_{ccb}^*B_c^{+}$ and $\Omega_{ccc}\eta_b$ channels
with a ratio of
\begin{eqnarray}
    \Gamma[\Omega_{ccb}^*B_c^{+}]:\Gamma[\Omega_{ccc}\eta_b]\simeq 3.
\end{eqnarray}
The partial widths of the other channels, such as $\Omega_{ccb}^{(*)}B_c^{*+}$,
are negligibly small. For the two $J^P=1/2^-$ states, the fall-apart widths
are comparable with that of the $J^P=3/2^-$ state.
The low-lying $J^P=1/2^-$ state $P_{c^3b\bar{b}}(14638)1/2^-$ mainly decays into the $\Omega_{ccb}B_c^{+}$ channel,
and also has sizeable decay rates into the $\Omega_{ccb}B_c^{*+}$
and $\Omega_{ccc}\Upsilon$ channels. The partial width ratios between these channels are estimated to be
\begin{eqnarray}
    \Gamma[\Omega_{ccb}B_c^{+}]:\Gamma[\Omega_{ccb}B_c^{*+}]:\Gamma[\Omega_{ccc}\Upsilon]\simeq 2.7:1.2:1.
\end{eqnarray}
While the high-lying $J^P=1/2^-$ state $P_{c^3b\bar{b}}(14664)1/2^-$ mainly decays into the $\Omega_{ccb}^*B_c^{*+}$ channel,
and also has sizeable decay rates into the $\Omega_{ccb}B_c^{*+}$
and $\Omega_{ccc}\Upsilon$ channels. The partial width ratios between
these main decay channels are predicted to be
\begin{eqnarray}
    \Gamma[\Omega_{ccb}^*B_c^{*+}]:\Gamma[\Omega_{ccb}B_c^{*+}]:\Gamma[\Omega_{ccc}\Upsilon]\simeq 4.4:1.4:1.
\end{eqnarray}
To establish these $1S$-wave $P_{c^3b\bar{b}}$ states, the $\Omega_{ccb}^{(*)}B_c^{+}$
and $\Omega_{ccb}^*B_c^{*+}$ channels are worth observing in experiments.

For the $bbbc\bar{c}$ system, the $J^P=3/2^-$ state $P_{b^3c\bar{c}}(17837)3/2^-$
may have a narrow fall-apart width of $\Gamma\simeq 0.56$ MeV,
and mainly decays into the $\Omega_{bbb}\eta_c$, $\Omega_{cbb}^*B_c^{-}$,
and $\Omega_{cbb}^*B_c^{*-}$ channels. The partial width ratios between these
channels are predicted to be
\begin{eqnarray}
    \Gamma[\Omega_{bbb}\eta_c]:\Gamma[\Omega_{cbb}^*B_c^{-}]:\Gamma[\Omega_{cbb}^*B_c^{*-}]\simeq 2.3:1:1.
\end{eqnarray}
Compared with the $J^P=3/2^-$ state, the low-lying $J^P=1/2^-$ state $P_{b^3c\bar{c}}(17824)1/2^-$
may have a smaller fall-apart width of $\Gamma\simeq 0.30$ MeV.
This state mainly decays into the $\Omega_{cbb}B_c^{-}$, $\Omega_{bbb}J/\psi$, and
and $\Omega_{cbb}^*B_c^{*+}$ channels. The partial width ratios between these
channels are predicted to be
\begin{eqnarray}
    \Gamma[\Omega_{cbb}B_c^{-}]:\Gamma[\Omega_{bbb}J/\psi]:\Gamma[\Omega_{cbb}^*B_c^{*-}]\simeq 2.5:1.5:1.
\end{eqnarray}
The high-lying $J^P=1/2^-$ state $P_{b^3c\bar{c}}(17856)1/2^-$
may have a fall-apart width of $\Gamma\simeq 0.99$ MeV,
and mainly decays into the $\Omega_{bbb}J/\psi$, $\Omega_{cbb}B_c^{-}$, and
and $\Omega_{cbb}^*B_c^{*+}$ channels. The partial width ratios between these
channels are predicted to be
\begin{eqnarray}
    \Gamma[\Omega_{bbb}J/\psi]:\Gamma[\Omega_{cbb}B_c^{-}]:\Gamma[\Omega_{cbb}^*B_c^{*-}]\simeq 3.0:1.3:1.
\end{eqnarray}
The above analysis indicates that the $\Omega_{bbb}\eta_c$, $\Omega_{bbb}J/\psi$,
and $\Omega_{cbb}B_c^{-}$ may be ideal channels for establishing the
$1S$-wave $P_{b^3c\bar{c}}$ states.

For the $bbbc\bar{b}$ system, the $J^P=3/2^-$ state $P_{b^3c\bar{b}}(21028)3/2^-$
may have a tiny fall-apart width of $\Gamma\simeq 0.41$ MeV,
and dominantly decays into the $\Omega_{bbb}B_c^+$ with a branching fraction of $\sim70\%$.
The fall-apart widths for the two $J^P=1/2^-$ states are comparable with that of the $J^P=3/2^-$ state.
The low-lying $J^P=1/2^-$ state $P_{b^3c\bar{b}}(21012)1/2^-$
mainly decays into the $\Omega_{bbb}B_c^{*+}$ and $\Omega_{cbb}\eta_b$ channels
with a partial width ratio
\begin{eqnarray}
    \Gamma[\Omega_{bbb}B_c^{*+}]:\Gamma[\Omega_{cbb}\eta_b]\simeq 1.5.
\end{eqnarray}
While the high-lying $J^P=1/2^-$ state $P_{b^3c\bar{b}}(21041)1/2^-$
dominantly decays into the $\Omega_{bbb}B_c^{*+}$ with a branching fraction $\sim69\%$.
The $\Omega_{cbb}\eta_b$ and $\Omega_{bbb}B_c^{+}$ may be ideal channels for establishing the
$1S$-wave $P_{b^3c\bar{b}}$ states.

\subsection{Pentaquarks with $\{12\}\{34\}\bar{5}$ symmetry}\label{12}

In the all-heavy pentaquarks, the $ccbb\bar{c}$ and $bbcc\bar{b}$
systems have the $\{12\}\{34\}\bar{5}$ symmetry. Considering the permutation symmetry
of identical quarks, for each pentaquark system there are nine $1S$ states,
one with $J^P=5/2^-$, four with $J^P=3/2^-$, and four with $J^P=1/2^-$.

\subsubsection{Mass}

The predicted mass spectrum has been given in Table~\ref{[123]45 Mixing}
and also shown in Fig.~\ref{aac}. For the $J^P=3/2^-$ and $1/2^-$ states,
there is sizeable mixing in the different color-spin configurations.

The masses of the $1S$ states for the $ccbb\bar{c}$ and $bbcc\bar{b}$ systems are predicted to
be $\sim14.6$ and $\sim17.8$ GeV, respectively.
These $1S$ states are far above the dissociation baryon-meson thresholds.
Some other studies about these pentaquarks can be found in the literature
based on the quark potential models~\cite{An:2022fvs,Yan:2023kqf,Gordillo:2024blx}, MIT bag model~\cite{Zhang:2023hmg},
CMI model~\cite{An:2020jix}, and effective mass and screened charge model~\cite{Rashmi:2024ako}.
There are large discrepancies (about $300-800$ MeV) in the predictions from various models.
Our predicted mass ranges for the $1S$ states are
close to those predicted in Ref.~\cite{An:2022fvs}.

The contributions from each part of the Hamiltonian for the pentaquark states
are given in Table~\ref{aa} as well.
Taking the $ccbb\bar{c}$ system as an example, from the table
one can see that the kinetic energy $\langle T \rangle\sim 920$ MeV, the confining
potential $\langle V^{Conf} \rangle\sim 690$ MeV, and the color-Coulomb potential
$\langle V^{Coul}\rangle\sim -1180$ MeV have the same order of magnitude.
By summing these contributions, $\langle T\rangle+\langle V^{Conf}\rangle+\langle V^{Coul}\rangle$,
for each $1S$ state, one obtains three nearly degenerate values 396, 415, and 443 MeV.
These values together with the small spin-spin potential $\langle V^{SS} \rangle$
determine the mass splittings between the different configurations.

In Table~\ref{aa} we further give the
root mean square (RMS) radii between two quarks. It is found that
the $1S$ states of the $ccbb\bar{c}$ and $bbcc\bar{b}$ systems are compact states,
the RMS radii are predicted to be in the range of $\sim0.28-0.52$ fm.

\subsubsection{Decay}

In Table~\ref{decayb}, the fall-apart decay properties of the $1S$ pentaquark
states are given. It is found that
the $1S$ states with $\{12\}\{34\}\bar{5}$ symmetry might be narrow states,
their fall-apart decay widths scatter in the range of $\sim0.1-1.9$ MeV.

For the $ccbb\bar{c}$ system, the $J^P=5/2^-$ state $P_{c^2b^2\bar{c}}(14619)5/2^-$
may have a fall-apart width of $\Gamma\simeq 1.06$ MeV,
which is dominantly contributed by the $\Omega_{ccb}^*B_c^{*-}$ channel. This state also
has a sizeable decay rate into the $\Omega_{cbb}^*J/\psi$ channel.
The partial width ratio between $\Omega_{cbb}^*J/\psi$ and $\Omega_{ccb}^*B_c^{*-}$ is estimated to be
\begin{eqnarray}
    \Gamma[\Omega_{cbb}^*J/\psi]:\Gamma[\Omega_{ccb}^*B_c^{*-}]\simeq 0.11.
\end{eqnarray}
The three low-lying $J^P=3/2^-$
states $P_{c^2b^2\bar{c}}(14544)3/2^-$, $P_{c^2b^2\bar{c}}(14575)3/2^-$, and $P_{c^2b^2\bar{c}}(14608)3/2^-$ have
relatively narrow fall-apart widths of 100s keV, and dominantly decay into the
$\Omega_{cbb}^*J/\psi$, $\Omega_{ccb}^*B_c^{*-}$, and $\Omega_{ccb}^*B_c^{-}$ channels, respectively.
The high-lying $J^P=3/2^-$ state $P_{c^2b^2\bar{c}}(14612)3/2^-$ may have a
broader fall-apart width of $\Gamma\simeq 1.72$ MeV. This state dominantly decays
into both the $\Omega_{ccb}B_c^{*-}$ and $\Omega_{ccb}^*B_c^{*-}$ channels.
The partial width ratio is predicted to be
\begin{eqnarray}
    \Gamma[\Omega_{ccb}B_c^{*-}]:\Gamma[\Omega_{ccb}^*B_c^{*-}]\simeq 0.9.
\end{eqnarray}
The low-lying $J^P=1/2^-$ state $P_{c^2b^2\bar{c}}(14540)1/2^-$ may have a
broader fall-apart width of $\Gamma\simeq 1.84$ MeV. This state dominantly decays
into the $\Omega_{ccb}B_c^{*-}$ channel, and also has sizeable decay
rates into the $\Omega_{ccb}^*B_c^{*-}$, $\Omega_{cbb}^*J/\psi$, and $\Omega_{ccb}B_c^{-}$ channels.
The partial width ratios are predicted to be
\begin{eqnarray}
    \Gamma[\Omega_{ccb}B_c^{*-}]:\Gamma[\Omega_{ccb}^*B_c^{*-}]:\Gamma[\Omega_{cbb}^*J/\psi]:
    \Gamma[\Omega_{ccb}B_c^{-}]\nonumber\\ \simeq 5.5:2.5:1.4:1.
\end{eqnarray}
The second $J^P=1/2^-$ state $P_{c^2b^2\bar{c}}(14572)1/2^-$ may have a very small fall-apart width of $\Gamma\simeq 0.19$ MeV, which
is mainly contributed by the $\Omega_{ccb}B_c^{-}$ and $\Omega_{ccb}^*B_c^{*-}$ channels.
The third $J^P=1/2^-$ state $P_{c^2b^2\bar{c}}(14609)1/2^-$ dominantly decays into the $\Omega_{ccb}B_c^{*-}$ channel
with a fall-apart width of $\Gamma\simeq 0.79$ MeV. The high-lying $J^P=1/2^-$
state $P_{c^2b^2\bar{c}}(14621)1/2^-$ has a fall-apart width of $\Gamma\simeq 0.51$ MeV. This state
mainly decays into the $\Omega_{ccb}B_c^{-}$, $\Omega_{ccb}B_c^{*-}$, and $\Omega_{cbb}J/\psi$ channels
with comparable decay rates. To establish the $1S$-wave $P_{c^2b^2\bar{c}}$ states,
the $\Omega_{ccb}^{(*)}B_c^{(*-)}$ and $\Omega_{cbb}^{(*)}J/\psi$ channels are worth observing in experiments.

For the $bbcc\bar{b}$ system, the $J^P=5/2^-$ state $P_{c^2b^2\bar{b}}(17824)5/2^-$
may have a fall-apart width of $\Gamma\simeq 0.82$ MeV,
which is dominantly contributed by the $\Omega_{cbb}^*B_c^{*+}$ channel. This state also
has a sizeable decay rate into the $\Omega_{ccb}^*\Upsilon$ channel.
The partial width ratio is estimated to be
\begin{eqnarray}
    \Gamma[\Omega_{ccb}^*\Upsilon]:\Gamma[\Omega_{cbb}^*B_c^{*+}] \simeq 0.21.
\end{eqnarray}
The low-lying $J^P=3/2^-$ state $P_{c^2b^2\bar{b}}(17759)3/2^-$ may have a fall-apart width of
$\Gamma\simeq 0.36$ MeV, which is mainly contributed by the $\Omega_{cbb}^*B_c^{+}$,
$\Omega_{ccb}\Upsilon$, $\Omega_{ccb}^*\Upsilon$ channels. The partial width ratios
between these channels are predicted to be
\begin{eqnarray}
    \Gamma[\Omega_{cbb}^*B_c^{+}]:\Gamma[\Omega_{ccb}^*\Upsilon]:\Gamma[\Omega_{ccb}\Upsilon]\simeq 1.6:1:1.
\end{eqnarray}
The second $J^P=3/2^-$ state $P_{c^2b^2\bar{b}}(17779)3/2^-$
has a very narrow fall-apart width of $\Gamma\simeq 0.15$ MeV, which is
mainly contributed by the $\Omega_{cbb}^*B_c^{*+}$,
$\Omega_{cbb}^*\eta_b$, $\Omega_{ccb}^*\Upsilon$ channels with ratios
\begin{eqnarray}
    \Gamma[\Omega_{cbb}^*B_c^{*+}]:\Gamma[\Omega_{cbb}^*\eta_b]:\Gamma[\Omega_{ccb}^*\Upsilon]\simeq 2:1:1.
\end{eqnarray}
The third $J^P=3/2^-$ state $P_{c^2b^2\bar{b}}(17815)3/2^-$
has a fall-apart width of $\Gamma\simeq 0.51$ MeV, which is
mainly contributed by the $\Omega_{cbb}^*B_c^{+}$ and $\Omega_{ccb}^*\Upsilon$
channels with a ratio
\begin{eqnarray}
    \Gamma[\Omega_{cbb}^*B_c^{+}]:\Gamma[\Omega_{ccb}^*\Upsilon]\simeq 2:1.
\end{eqnarray}
The high-lying $J^P=3/2^-$ state $P_{c^2b^2\bar{b}}(17818)3/2^-$ has a fall-apart width
of $\Gamma\simeq 0.63$ MeV, and mainly decay into both the $\Omega_{cbb}B_c^{*+}$ and
$\Omega_{cbb}^*B_c^{*+}$ channels with comparable decay rates, i.e.,
\begin{eqnarray}
    \Gamma[\Omega_{cbb}B_c^{*+}]:\Gamma[\Omega_{cbb}^*B_c^{*+}]\simeq 1.1.
\end{eqnarray}
The low-lying $J^P=1/2^-$ state $P_{c^2b^2\bar{b}}(17761)1/2^-$ may have a fall-apart width
of $\Gamma\simeq 0.49$ MeV, which is mainly contributed by
the $\Omega_{cbb}B_c^{*+}$, $\Omega_{cbb}^*B_c^{*+}$, and $\Omega_{ccb}^*\Upsilon$ channels.
The partial width ratios are predicted to be
\begin{eqnarray}
    \Gamma[\Omega_{cbb}B_c^{*+}]:\Gamma[\Omega_{cbb}^*B_c^{*+}]:\Gamma[\Omega_{ccb}^*\Upsilon]\simeq 1.2:1:1.
\end{eqnarray}
The high-lying three $J^P=1/2^-$ states
$P_{c^2b^2\bar{b}}(17770)1/2^-$, $P_{c^2b^2\bar{b}}(17813)1/2^-$, and $P_{c^2b^2\bar{b}}(17824)1/2^-$ have a
comparable fall-apart decay width of $\Gamma\sim 300$ keV, and dominantly
decay into the $\Omega_{ccb}^*\Upsilon$, $\Omega_{cbb}B_c^{*+}$, and $\Omega_{cbb}B_c^{+}$ channels, respectively.
To look for the $1S$-wave $P_{c^2b^2\bar{b}}$ states,
the $\Omega_{cbb}^{(*)}B_c^{(*+)}$ and $\Omega_{ccb}^{(*)}\Upsilon$ channels are worth observing in experiments.

\section{Summary}\label{sum}

In this work, we carry out a dynamical calculation of the mass spectra for the all-heavy pentaquarks with a
non-relativistic potential model. To precisely treat a five-body system, we apply the ECG method, in which a variational trial
spatial wave function is expanded with nondiagonal Gaussians associated to the Jacobi coordinates.
A complete mass spectrum for the $1S$ states is obtained. The masses of the $1S$-wave $P_{c^4\bar{c}}$, $P_{c^4\bar{b}}$,
$P_{b^4\bar{c}}$, and $P_{b^4\bar{b}}$ states with the $\{1234\}\bar{5}$ symmetry
are predicted to be in the ranges of $\sim8.2$, $\sim11.5$, $\sim21.1$ and $\sim24.2$ GeV, respectively.
The masses of the $1S$-wave $P_{c^3b\bar{c}}$, $P_{c^3b\bar{b}}$,
$P_{b^3c\bar{c}}$, and $P_{b^3c\bar{b}}$ states with the $\{123\}4\bar{5}$ symmetry
are predicted to be in the ranges of $\sim11.4$, $\sim14.6$,
$\sim17.8$, and $\sim21.0$ GeV, respectively. While for the $P_{c^2b^2\bar{c}}$ and $P_{c^2b^2\bar{b}}$ with $\{12\}\{34\}\bar{5}$ symmetry,
the masses of the $1S$ states are predicted to be in the range of $\sim14.6$ and $\sim17.8$ GeV, respectively.
The obtained pentaquark states should have a compact structure, and lie far above the lowest
dissociation baryon-meson mass threshold.

Moreover, by using the masses and wave functions obtained from the potential model we further evaluate the
the fall-apart decay properties of the $1S$ pentaquark states within a
quark-exchange model. The all-heavy pentaquarks may have narrow fall-apart decay widths,
which scatter in the range of $\sim0.1-4.0$ MeV. The main fall-apart decay
channels and the partial width ratios are given, which may provide useful references for
future experimental searches.

\section*{Acknowledgements}

We thank Yan-Rui Liu for part contributions to this work. We also thank Hong-Tao An for useful discussions
about the pentaquark configurations. This work is supported by the National Natural
Science Foundation of China under Grants Nos.12175065 and 12235018.


\section*{APPENDIX}\label{APPENDIX}

The Young tableau can be represented as $[f_{1}f_{2}\cdots f_{m}]_{r_{n}r_{n-1}\cdots}$, where $f_{1},f_{2},\cdots,f_{m}$
denote the number of grids of the first, second, $\cdots$, and $m$-th rows of the Young diagram, respectively.
The subscripts $r_{n},r_{n-1},\cdots$ indicate the numbers $n$, $n-1$, $\cdots$, appearing
in the $r_n$, $r_{n-1}$-th, $\cdots$, rows, respectively. For example, the
representations $[31]_{2}$, $[31]_{12}$, $[31]_{11}$ of the $S_4$ group corresponding to the Young tableaux
\begin{tabular}{|c|c|c|}
	\hline
	1 & 2 & 3\\
	\hline
	4 & \multicolumn{1}{c}{} & \multicolumn{1}{c}{}\\
	\cline{1-1}
\end{tabular},
\begin{tabular}{|c|c|c|}
	\hline
	1 & 2 & 4\\
	\hline
	3 & \multicolumn{1}{c}{} & \multicolumn{1}{c}{}\\
	\cline{1-1}
\end{tabular}, and
\begin{tabular}{|c|c|c|}
	\hline
	1 & 3 & 4\\
	\hline
	2 & \multicolumn{1}{c}{} & \multicolumn{1}{c}{}\\
	\cline{1-1}
\end{tabular}, respectively.


First, we give an example for constructing the color-spin wave functions of the pentaquark system with $\{1234\}\bar{5}$ symmetry.
For the antisymmetry requirement of the four identical quarks $\{1234\}$, the representation
of color$\otimes$spin space should be $[1^4]_4^{CS}$, which can be expressed by~\cite{Stancu:1999qr}
\begin{eqnarray}
	\label{5}
	[1^{4}]_{4}^{CS}  =  \sqrt{\frac{1}{3}}[31]_{2}^{S}[211]_{1}^{C}-\sqrt{\frac{2}{3}}[31]_{1}^{S}\otimes[211]_{3}^{C}.
\end{eqnarray}
It should be emphasized that the representations $[31]_{1}$=%
\begin{tabular}{|c|c|c|}
	\hline
	1 & $\;$ & 4\\
	\hline
	& \multicolumn{1}{c}{} & \multicolumn{1}{c}{}\\
	\cline{1-1}
\end{tabular} and $[211]_{3}$=%
\begin{tabular}{|c|c}
	\hline
	1 & \multicolumn{1}{c|}{$\;$}\\
	\hline
	& \\
	\cline{1-1}
	4 & \\
	\cline{1-1}
\end{tabular} in Eq.~(\ref{5}) do not correspond to any specific Young tableaux.
The symmetry of the subsystem $\{123\}$ should be further determined. By the coupling of $[21]^S\otimes[21]^C$ in the $S_3$ group,
the representation of the $\{123\}$ should be $[1^{3}]_{3}^{CS}$:
\begin{eqnarray}
	\label{6}
	[1^{3}]_{3}^{CS}  =  \sqrt{\frac{1}{2}}[21]_{2}^{S}[21]_{1}^{C}-\sqrt{\frac{1}{2}}[21]_{1}^{S}[21]_{2}^{C}.
\end{eqnarray}
Combining Eq.~(\ref{5}) and Eq.~(\ref{6}), we easily obtain
\begin{widetext}
\begin{eqnarray}
	[1^{4}]_4^{CS} = & \sqrt{\frac{1}{3}}[31]_{2}^{S}[211]_{1}^{C}-\sqrt{\frac{2}{3}}\left(\sqrt{\frac{1}{2}}[31]_{12}^{S}[211]_{31}^{C}-\sqrt{\frac{1}{2}}[31]_{11}^{S}[211]_{32}^{C}\right) \\
	              = & \sqrt{\frac{1}{3}}[31]_{2}^{S}[211]_{1}^{C}-\sqrt{\frac{1}{3}}[31]_{12}^{S}[211]_{31}^{C}+\sqrt{\frac{1}{3}}[31]_{11}^{S}[211]_{32}^{C}.\nonumber
\end{eqnarray}
\end{widetext}
Expressing it with the Young tableaux, one has
\begin{widetext}
	\begin{eqnarray}
		\begin{aligned}
			\begin{tabular}{|c|}
				\hline
				1\\
				\hline
				2\\
				\hline
				3\\
				\hline
				4\\
				\hline
			\end{tabular}^{CS}=\sqrt{\frac{1}{3}}%
			\begin{tabular}{|c|c}
				\hline
				1 & \multicolumn{1}{c|}{4}\\
				\hline
				2 & \\
				\cline{1-1}
				3 & \\
				\cline{1-1}
			\end{tabular}^{C}%
			\begin{tabular}{|c|c|c|}
				\hline
				1 & 2 & 3\\
				\hline
				4 & \multicolumn{1}{c}{} & \multicolumn{1}{c}{}\\
				\cline{1-1}
			\end{tabular}^{S}+\sqrt{\frac{1}{3}}%
			\begin{tabular}{|c|c}
				\hline
				1 & \multicolumn{1}{c|}{2}\\
				\hline
				3 & \\
				\cline{1-1}
				4 & \\
				\cline{1-1}
			\end{tabular}^{C}%
			\begin{tabular}{|c|c|c|}
				\hline
				1 & 3 & 4\\
				\hline
				2 & \multicolumn{1}{c}{} & \multicolumn{1}{c}{}\\
				\cline{1-1}
			\end{tabular}^{S}-\sqrt{\frac{1}{3}}%
			\begin{tabular}{|c|c}
				\hline
				1 & \multicolumn{1}{c|}{3}\\
				\hline
				2 & \\
				\cline{1-1}
				4 & \\
				\cline{1-1}
			\end{tabular}^{C}%
			\begin{tabular}{|c|c|c|}
				\hline
				1 & 2 & 4\\
				\hline
				3 & \multicolumn{1}{c}{} & \multicolumn{1}{c}{}\\
				\cline{1-1}
			\end{tabular}^{S}.
		\end{aligned}
	\end{eqnarray}
\end{widetext}
Finally, the fifth antiquark should be included. Note that the three color configurations given in Eq.~(\ref{color})
can be further expressed as
\begin{eqnarray}
	\begin{aligned}
		\label{onebyonecolor}
C_1=\begin{tabular}{|c|c|}
	\hline
	1 & 4\\
	\hline
	2 & $\alpha$\\
	\hline
	3 & $\beta$\\
	\hline
\end{tabular},\;
C_2=\begin{tabular}{|c|c|}
	\hline
	1 & 2\\
	\hline
	3 & $\alpha$\\
	\hline
	4 & $\beta$\\
	\hline
\end{tabular},\;
C_3=\begin{tabular}{|c|c|}
	\hline
	1 & 3\\
	\hline
	2 & $\alpha$\\
	\hline
	4 & $\beta$\\
	\hline
\end{tabular}.
	\end{aligned}
\end{eqnarray}
By combining the spin configurations given in Eqs.~(\ref{a32}) and~(\ref{a12}),
we get the color-spin wave functions for the $J^P=3/2^-$ and $1/2^-$ states as follows:
\begin{widetext}
	\begin{eqnarray}
		\begin{aligned}
			\left|\begin{tabular}{|c|}
				\hline
				1\\
				\hline
				2\\
				\hline
				3\\
				\hline
				4\\
				\hline
			\end{tabular}\otimes\bar{5}\right\rangle^{CS}_{3/2}&=\sqrt{\frac{1}{3}}%
			\begin{tabular}{|c|c|}
				\hline
				1 & 4 \\
				\hline
				2 & $\alpha$ \\
				\hline
				3 & $\beta$ \\
				\hline
			\end{tabular}^{C}\text{\ensuremath{\;}}%
			\begin{tabular}{|c|c|c|c}
				\cline{1-3} \cline{2-3} \cline{3-3}
				1 & 2 & 3 & 5\\
				\cline{1-3} \cline{2-3} \cline{3-3}
				4 & \multicolumn{1}{c}{} & \multicolumn{1}{c}{} & \\
				\cline{1-1}
			\end{tabular}^{S}+\sqrt{\frac{1}{3}}%
			\begin{tabular}{|c|c|}
				\hline
				1 & 2\\
				\hline
				3 & $\alpha$\\
				\hline
				4 & $\beta$\\
				\hline
			\end{tabular}^{C}\text{\ensuremath{\;}}%
			\begin{tabular}{|c|c|c|c}
				\cline{1-3} \cline{2-3} \cline{3-3}
				1 & 3 & 4 & 5\\
				\cline{1-3} \cline{2-3} \cline{3-3}
				2 & \multicolumn{1}{c}{} & \multicolumn{1}{c}{} & \\
				\cline{1-1}
			\end{tabular}^{S}-\sqrt{\frac{1}{3}}%
			\begin{tabular}{|c|c|}
				\hline
				1 & 3\\
				\hline
				2 & $\alpha$\\
				\hline
				4 & $\beta$\\
				\hline
			\end{tabular}^{C}\text{\ensuremath{\;}}%
			\begin{tabular}{|c|c|c|c}
				\cline{1-3} \cline{2-3} \cline{3-3}
				1 & 2 & 4 & 5\\
				\cline{1-3} \cline{2-3} \cline{3-3}
				3 & \multicolumn{1}{c}{} & \multicolumn{1}{c}{} & \\
				\cline{1-1}
			\end{tabular}^{S},\\
			\left|\begin{tabular}{|c|}
				\hline
				1\\
				\hline
				2\\
				\hline
				3\\
				\hline
				4\\
				\hline
			\end{tabular}\otimes\bar{5}\right\rangle^{CS}_{1/2}&=\sqrt{\frac{1}{3}}%
			\begin{tabular}{|c|c|}
				\hline
				1 & 4\\
				\hline
				2 & $\alpha$\\
				\hline
				3 & $\beta$\\
				\hline
			\end{tabular}^{C}\;%
			\begin{tabular}{|c|c|c|}
				\hline
				1 & 2 & 3\\
				\hline
				4 & \multicolumn{1}{c}{5} & \multicolumn{1}{c}{}\\
				\cline{1-1}
			\end{tabular}^{S}+\sqrt{\frac{1}{3}}%
			\begin{tabular}{|c|c|}
				\hline
				1 & 2\\
				\hline
				3 & $\alpha$\\
				\hline
				4 & $\beta$\\
				\hline
			\end{tabular}^{C}\;%
			\begin{tabular}{|c|c|c|}
				\hline
				1 & 3 & 4\\
				\hline
				2 & \multicolumn{1}{c}{5} & \multicolumn{1}{c}{}\\
				\cline{1-1}
			\end{tabular}^{S}-\sqrt{\frac{1}{3}}%
			\begin{tabular}{|c|c|}
				\hline
				1 & 3\\
				\hline
				2 & $\alpha$\\
				\hline
				4 & $\beta$\\
				\hline
			\end{tabular}^{C}\;%
			\begin{tabular}{|c|c|c|}
				\hline
				1 & 2 & 4\\
				\hline
				3 & \multicolumn{1}{c}{5} & \multicolumn{1}{c}{}\\
				\cline{1-1}
			\end{tabular}^{S},
		\end{aligned}
	\end{eqnarray}
\end{widetext}
which just correspond to the two configurations given in Eqs.~(\ref{1234}) and (\ref{1234b}).


While for the pentaquark system with $\{123\}4\bar{5}$ symmetry, we take the
$1S_{\frac{3}{2}^-}(\{123\}4\bar{5})_3$ configuration as an example for constructing the color-spin wave function.
For the antisymmetry requirement of the three identical quarks $\{123\}$, the representation
of color$\otimes$spin space should be $[1^3]_3^{CS}$, which has been given by Eq.~(\ref{6}).
Expressing it with the Young tableaux, one has
\begin{eqnarray}
	\begin{aligned}
		\begin{tabular}{|c|}
			\hline
			1\\
			\hline
			2\\
			\hline
			3\\
			\hline
		\end{tabular}^{CS}=\sqrt{\frac{1}{2}}%
		\begin{tabular}{|c|c}
			\hline
			1 & \multicolumn{1}{c|}{3}\\
			\hline
			2 & \\
			\cline{1-1}
		\end{tabular}^{C}%
		\begin{tabular}{|c|c|}
			\hline
			1 & 2\\
			\hline
			3 & \multicolumn{1}{c}{} \\
			\cline{1-1}
		\end{tabular}^{S}-\sqrt{\frac{1}{2}}%
		\begin{tabular}{|c|c}
			\hline
			1 & \multicolumn{1}{c|}{2}\\
			\hline
			3 & \\
			\cline{1-1}
		\end{tabular}^{C}%
		\begin{tabular}{|c|c|}
			\hline
			1 & 3\\
			\hline
			2 & \multicolumn{1}{c}{}\\
			\cline{1-1}
		\end{tabular}^{S}
	\end{aligned}
\end{eqnarray}
Finally, the fourth quark and the fifth antiquark should be included. By combining the spin configurations given in Eq.~(\ref{a32})
and the color configurations given in Eq.~(\ref{onebyonecolor}),
we get the color-spin wave functions for the $1S_{\frac{3}{2}^-}(\{123\}4\bar{5})_3$ configuration,
\begin{widetext}
\begin{eqnarray}
	\begin{aligned}
		\left|\begin{tabular}{|c|}
			\hline
			1\\
			\hline
			2\\
			\hline
			3\\
			\hline
		\end{tabular}\otimes4\otimes\bar{5}\right\rangle^{CS}_{3/2}&=\sqrt{\frac{1}{2}}%
		\begin{tabular}{|c|c|}
			\hline
			1 & 3 \\
			\hline
			2 & $\alpha$ \\
			\hline
			4 & $\beta$ \\
			\hline
		\end{tabular}^{C}\text{\ensuremath{\;}}%
		\begin{tabular}{|c|c|c|c}
			\cline{1-3} \cline{2-3} \cline{3-3}
			1 & 2 & 4 & 5\\
			\cline{1-3} \cline{2-3} \cline{3-3}
			3 & \multicolumn{1}{c}{} & \multicolumn{1}{c}{} & \\
			\cline{1-1}
		\end{tabular}^{S}-\sqrt{\frac{1}{2}}%
		\begin{tabular}{|c|c|}
			\hline
			1 & 2\\
			\hline
			3 & $\alpha$\\
			\hline
			4 & $\beta$\\
			\hline
		\end{tabular}^{C}\text{\ensuremath{\;}}%
		\begin{tabular}{|c|c|c|c}
			\cline{1-3} \cline{2-3} \cline{3-3}
			1 & 3 & 4 & 5\\
			\cline{1-3} \cline{2-3} \cline{3-3}
			2 & \multicolumn{1}{c}{} & \multicolumn{1}{c}{} & \\
			\cline{1-1}
		\end{tabular}^{S}
	\end{aligned}
\end{eqnarray}
\end{widetext}
which just correspond to the configuration given in Eq.~(\ref{equation 123b}).


For the pentaquark system with symmetry $\{12\}\{34\}\bar{5}$, the details
for constructing the spin-color wave functions are given as follows. 
The identical quark pairs $(Q_1Q_2)$ and $(Q_3Q_4)$ should satisfy the requirements 
of the $S_2$ group, respectively. 
For convenience, first we construct the color and spin wave functions by using the
diquark-diquark-antiquark form, respectively.

For this form, the color wave functions are given by
\begin{eqnarray}
		C_1'&=[(\{12\}^{\bar{3}}\{34\}^{\bar{3}})^3\bar{5}^{\bar{3}}]^1 \nonumber,\\
		C_2'&=[(\{12\}^6\{34\}^{\bar{3}})^3\bar{5}^{\bar{3}}]^1,\\
		C_3'&=[(\{12\}^{\bar{3}}\{34\}^6)^3\bar{5}^{\bar{3}}]^1,\nonumber
\end{eqnarray}
where, the superscripts are the color representations of SU(3) group.
On the other hand, the Young tableaux representations of the color wave functions are given by
\begin{eqnarray}
	\label{diquarkdiantiColor}
	C_1'=&\begin{tabular}{|c|}
		\hline
		1\tabularnewline
		\hline
		2\tabularnewline
		\hline
	\end{tabular}\otimes%
	\begin{tabular}{|c|}
		\hline
		3\tabularnewline
		\hline
		4\tabularnewline
		\hline
	\end{tabular}\otimes%
	\begin{tabular}{|c|}
		\hline
		$\alpha$\tabularnewline
		\hline
		$\beta$\tabularnewline
		\hline
	\end{tabular}
	&=\begin{tabular}{|c|c|}
		\hline
		1 & 3\tabularnewline
		\hline
		2 & $\alpha$\tabularnewline
		\hline
		4 & $\beta$\tabularnewline
		\hline
	\end{tabular} \nonumber,\\
	C_2'=&\begin{tabular}{|c|c|}
		\hline
		1 & 2\tabularnewline
		\hline
	\end{tabular}\otimes%
	\begin{tabular}{|c|}
		\hline
		3\tabularnewline
		\hline
		4\tabularnewline
		\hline
	\end{tabular}\otimes%
	\begin{tabular}{|c|}
		\hline
		$\alpha$\tabularnewline
		\hline
		$\beta$\tabularnewline
		\hline
	\end{tabular}
	&=\begin{tabular}{|c|c|}
		\hline
		1 & 2\tabularnewline
		\hline
		3 & $\alpha$\tabularnewline
		\hline
		4 & $\beta$\tabularnewline
		\hline
	\end{tabular},\\
	C_3'=&\begin{tabular}{|c|}
		\hline
		1\tabularnewline
		\hline
		2\tabularnewline
		\hline
	\end{tabular}\otimes%
	\begin{tabular}{|c|c|}
		\hline
		3 & 4\tabularnewline
		\hline
	\end{tabular}\otimes%
	\begin{tabular}{|c|}
		\hline
		$\alpha$\tabularnewline
		\hline
		$\beta$\tabularnewline
		\hline
	\end{tabular}
	&=\begin{tabular}{|c|c|}
		\hline
		1 & 4\tabularnewline
		\hline
		2 & $\alpha$\tabularnewline
		\hline
		3 & $\beta$\tabularnewline
		\hline
	\end{tabular}\nonumber.
\end{eqnarray}
It should be noted that the color wave functions expressed with
the diquark-diquark-antiquark form here are different from those of baryon-meson form given by Eq.~(\ref{onebyonecolor}).
The color wave functions of the two different forms can be related by
\begin{eqnarray}\label{ColorT}
	\begin{pmatrix}
		C_1' \\
		C_2' \\
		C_3'
	\end{pmatrix}
	=
	\begin{pmatrix}
		\sqrt{\frac{1}{3}}  & 0 & -\sqrt{\frac{2}{3}} \\
		0                   & 1 & 0                   \\
		-\sqrt{\frac{2}{3}} & 0 & -\sqrt{\frac{1}{3}}
	\end{pmatrix}
	\begin{pmatrix}
		C_1 \\
		C_2 \\
		C_3
	\end{pmatrix}.
\end{eqnarray}

Similarly, the spin wave functions with the diquark-diquark-antiquark structure are constructed as follows.

For $S=5/2$:
\begin{eqnarray}
		S_1' & =[(\{12\}_1\{34\}_1)_2\bar{5}_{1/2}]_{5/2}.
\end{eqnarray}

For $S=3/2$:
\begin{eqnarray}
	\begin{aligned}
		S_2' & =[(\{12\}_1\{34\}_1)_2\bar{5}_{1/2}]_{3/2}, \\
		S_3' & =[(\{12\}_1\{34\}_1)_1\bar{5}_{1/2}]_{3/2}, \\
		S_4' & =[(\{12\}_0\{34\}_1)_1\bar{5}_{1/2}]_{3/2}, \\
		S_5' & =[(\{12\}_1\{34\}_0)_1\bar{5}_{1/2}]_{3/2}.
	\end{aligned}
\end{eqnarray}

For $S=1/2$:
\begin{eqnarray}
	\begin{aligned}
		 S_6'   & =[(\{12\}_1\{34\}_1)_1\bar{5}_{1/2}]_{1/2}, \\
		 S_7'   & =[(\{12\}_0\{34\}_1)_1\bar{5}_{1/2}]_{1/2}, \\
		 S_8'   & =[(\{12\}_1\{34\}_0)_1\bar{5}_{1/2}]_{1/2}, \\
		 S_9'   & =[(\{12\}_1\{34\}_1)_0\bar{5}_{1/2}]_{1/2}, \\
		S_{10}' & =[(\{12\}_0\{34\}_0)_0\bar{5}_{1/2}]_{1/2}.
	\end{aligned}
\end{eqnarray}
The subscripts in the spin configuration represent the spin quantum numbers.
Their representations with Young tableaux are given as follows.

For $S=5/2$:
\begin{eqnarray}
S_1'=\begin{tabular}{|c|c|}
	\hline
	1 & 2 \tabularnewline \hline
\end{tabular}\otimes%
\begin{tabular}{|c|c|}
	\hline
	3 & 4 \tabularnewline \hline
\end{tabular}\otimes%
\begin{tabular}{|c|}
	\hline
	5 \tabularnewline \hline
\end{tabular}
=\begin{tabular}{|c|c|c|c|c|}
	\hline
	1 & 2 & 3 & 4 & 5 \tabularnewline \hline
\end{tabular}.
\end{eqnarray}

For $S=3/2$:
\begin{eqnarray}
	\begin{aligned}
		S_2'=&\begin{tabular}{|c|c|}
		\hline
		1 & 2 \tabularnewline \hline
	\end{tabular}\otimes%
	\begin{tabular}{|c|c|}
		\hline
		3 & 4 \tabularnewline \hline
	\end{tabular}\otimes%
	\begin{tabular}{|c|}
		\hline
		5 \tabularnewline \hline
	\end{tabular}
	&=\begin{tabular}{|c|c|c|c|}
		\hline
		1 & 2 & 3 & 4\tabularnewline
		\hline
		5 & \multicolumn{1}{c}{} & \multicolumn{1}{c}{} & \multicolumn{1}{c}{}\tabularnewline
		\cline{1-1}
	\end{tabular},\\
	S_3'=&\begin{tabular}{|c|c|}
		\hline
		1 & 2 \tabularnewline \hline
	\end{tabular}\otimes%
	\begin{tabular}{|c|c|}
		\hline
		3 & 4 \tabularnewline \hline
	\end{tabular}\otimes%
	\begin{tabular}{|c|}
		\hline
		5 \tabularnewline \hline
	\end{tabular}
	&=\begin{tabular}{|c|c|c|c|}
		\hline
		1 & 2 & 4 & 5\tabularnewline
		\hline
		3 & \multicolumn{1}{c}{} & \multicolumn{1}{c}{} & \multicolumn{1}{c}{}\tabularnewline
		\cline{1-1}
	\end{tabular},\\
	S_4'=&\begin{tabular}{|c|}
		\hline
		1 \tabularnewline \hline
		2 \tabularnewline \hline
	\end{tabular}\otimes%
	\begin{tabular}{|c|c|}
		\hline
		3 & 4 \tabularnewline \hline
	\end{tabular}\otimes%
	\begin{tabular}{|c|}
		\hline
		5 \tabularnewline \hline
	\end{tabular}
	&=\begin{tabular}{|c|c|c|c|}
		\hline
		1 & 3 & 4 & 5\tabularnewline
		\hline
		2 & \multicolumn{1}{c}{} & \multicolumn{1}{c}{} & \multicolumn{1}{c}{}\tabularnewline
		\cline{1-1}
	\end{tabular},\\
	S_5'=&\begin{tabular}{|c|c|}
		\hline
		1 & 2 \tabularnewline \hline
	\end{tabular}\otimes%
	\begin{tabular}{|c|}
		\hline
		3 \tabularnewline \hline
		4 \tabularnewline \hline
	\end{tabular}\otimes%
	\begin{tabular}{|c|}
		\hline
		5 \tabularnewline \hline
	\end{tabular}
	&=\begin{tabular}{|c|c|c|c|}
		\hline
		1 &          2           &          3           &          5           \tabularnewline \hline
		4 & \multicolumn{1}{c}{} & \multicolumn{1}{c}{} & \multicolumn{1}{c}{} \tabularnewline \cline{1-1}
	\end{tabular}.\\
	\end{aligned}
\end{eqnarray}

For $S=1/2$:
\begin{eqnarray}
	\begin{aligned}
		S_6'=&\begin{tabular}{|c|c|}
			\hline
			1 & 2 \tabularnewline \hline
		\end{tabular}\otimes%
		\begin{tabular}{|c|c|}
			\hline
			3 & 4 \tabularnewline \hline
		\end{tabular}\otimes%
		\begin{tabular}{|c|}
			\hline
			5 \tabularnewline \hline
		\end{tabular}
		&=\begin{tabular}{|c|c|c|}
			\hline
			1 & 2 &          5           \tabularnewline \hline
			3 & 4 & \multicolumn{1}{c}{} \tabularnewline \cline{1-2}
		\end{tabular},\\
		S_7'=&\begin{tabular}{|c|}
			\hline
			1 \tabularnewline \hline
			2 \tabularnewline \hline
		\end{tabular}\otimes%
		\begin{tabular}{|c|c|}
			\hline
			3 & 4 \tabularnewline \hline
		\end{tabular}\otimes%
		\begin{tabular}{|c|}
			\hline
			5 \tabularnewline \hline
		\end{tabular}
		&=\begin{tabular}{|c|c|c|}
			\hline
			1 & 4 & 5 \tabularnewline
			\hline
			2 & 3 & \multicolumn{1}{c}{} \tabularnewline
			\cline{1-2}
		\end{tabular},\\
		S_8'=&\begin{tabular}{|c|c|}
			\hline
			1 & 2 \tabularnewline \hline
		\end{tabular}\otimes%
		\begin{tabular}{|c|}
			\hline
			3 \tabularnewline \hline
			4 \tabularnewline \hline
		\end{tabular}\otimes%
		\begin{tabular}{|c|}
			\hline
			5 \tabularnewline \hline
		\end{tabular}
		&=\begin{tabular}{|c|c|c|}
			\hline
			1 & 2 &          3           \tabularnewline \hline
			4 & 5 & \multicolumn{1}{c}{} \tabularnewline \cline{1-2}
		\end{tabular},\\
		S_9'=&\begin{tabular}{|c|c|}
			\hline
			1 & 2 \tabularnewline \hline
		\end{tabular}\otimes%
		\begin{tabular}{|c|c|}
			\hline
			3 & 4 \tabularnewline \hline
		\end{tabular}\otimes%
		\begin{tabular}{|c|}
			\hline
			5 \tabularnewline \hline
		\end{tabular}
		&=\begin{tabular}{|c|c|c|}
			\hline
			1 & 2 &          4           \tabularnewline \hline
			3 & 5 & \multicolumn{1}{c}{} \tabularnewline \cline{1-2}
		\end{tabular},\\
		S_{10}'=&\begin{tabular}{|c|}
			\hline
			1 \tabularnewline \hline 2 \tabularnewline \hline
		\end{tabular}\otimes%
		\begin{tabular}{|c|}
			\hline
			3 \tabularnewline \hline 4 \tabularnewline \hline
		\end{tabular}\otimes%
		\begin{tabular}{|c|}
			\hline
			5 \tabularnewline \hline
		\end{tabular}
		&=\begin{tabular}{|c|c|c|}
			\hline
			1 & 3 &          5           \tabularnewline \hline
			2 & 4 & \multicolumn{1}{c}{} \tabularnewline \cline{1-2}
		\end{tabular}.
	\end{aligned}
\end{eqnarray}

The spin wave functions of the diquark-diquark-antiquark form can be can be related
to those given in Eqs.~(\ref{a82})-(\ref{a12}) by
\begin{eqnarray}\label{Spin1}
	S_1'=S_1,
\end{eqnarray}
\begin{eqnarray}\label{Spin2}
	\begin{pmatrix}
		S_2' \\
		S_3' \\
		S_4' \\
		S_5'
	\end{pmatrix}
	=
	\begin{pmatrix}
		1 & 0                  & 0 & 0                   \\
		0 & \sqrt{\frac{1}{3}} & 0 & \sqrt{\frac{2}{3}}  \\
		0 & 0                  & 1 & 0                   \\
		0 & \sqrt{\frac{2}{3}} & 0 & -\sqrt{\frac{1}{3}}
	\end{pmatrix}
	\begin{pmatrix}
		S_2 \\
		S_3 \\
		S_4 \\
		S_5
	\end{pmatrix},
\end{eqnarray}
\begin{eqnarray}\label{Spin3}
	\begin{pmatrix}
		S_6'    \\
		S_7'    \\
		S_8'    \\
		S_9'    \\
		S_{10}'
	\end{pmatrix}
	=
\begin{pmatrix}
	\sqrt{\frac{1}{3}} & 0 & \sqrt{\frac{2}{3}}  & 0 & 0 \\
	0                  & 1 & 0                   & 0 & 0 \\
	\sqrt{\frac{2}{3}} & 0 & -\sqrt{\frac{1}{3}} & 0 & 0 \\
	0                  & 0 & 0                   & 1 & 0 \\
	0                  & 0 & 0                   & 0 & 1
\end{pmatrix}
	\begin{pmatrix}
		S_6    \\
		S_7    \\
		S_8    \\
		S_9    \\
		S_{10}
	\end{pmatrix}.
\end{eqnarray}

In the color and spin wave functions constructed
by the diquark-diquark-antiquark form, there is explicit
permutation symmetry for the identical quark pairs $(Q_1Q_2)$ and $(Q_3Q_4)$,
which makes it easy to construct the color-spin wave functions of the pentaquarks.
For example, if the color symmetry for the $(Q_1Q_2)$ or $(Q_3Q_4)$
in the wave function is symmetric (antisymmetric), then the spin symmetry
for the $(Q_1Q_2)$ or $(Q_3Q_4)$ should be antisymmetric (symmetric).
The color-spin configurations constructed by the diquark-diquark-antiquark form are given as follows.
	
For $J^P=\frac{5}{2}^-$:
	\begin{eqnarray}
		1S_{\frac{5}{2}^-}(\{12\}\{34\}\bar{5})=C_1'S_1'.
	\end{eqnarray}
	
For $J^P=\frac{3}{2}^-$:
	\begin{equation}
		\begin{array}{cc}
			1S_{\frac{3}{2}^-}(\{12\}\{34\}\bar{5})_1&=C_1'S_2',\\
			1S_{\frac{3}{2}^-}(\{12\}\{34\}\bar{5})_2&=C_1'S_3',\\
			1S_{\frac{3}{2}^-}(\{12\}\{34\}\bar{5})_3&=C_2'S_4',\\
			1S_{\frac{3}{2}^-}(\{12\}\{34\}\bar{5})_4&=C_3'S_5',\\
		\end{array}
	\end{equation}
	
For $J^P=\frac{1}{2}^-$:
	\begin{equation}
		\begin{array}{cc}
			1S_{\frac{1}{2}^-}(\{12\}\{34\}\bar{5})_1&=C_1'S_6',\\
			1S_{\frac{1}{2}^-}(\{12\}\{34\}\bar{5})_2&=C_2'S_7',\\
			1S_{\frac{1}{2}^-}(\{12\}\{34\}\bar{5})_3&=C_3'S_8',\\
			1S_{\frac{1}{2}^-}(\{12\}\{34\}\bar{5})_4&=C_1'S_9'.\\
		\end{array}
	\end{equation}
	
By using the transformation relations given in Eq.~(\ref{ColorT}) for the color configurations,
and the relations given in Eqs.~(\ref{Spin1})-(\ref{Spin3}) for the spin configurations, one
can easily transform the spin-color wave functions constructed by the
diquark-diquark-antiquark form into the baryon-meson form given by the
Eqs.~(\ref{eq 2c5/2})-(\ref{eq 2c1/2}).



\bibliographystyle{unsrt}

\begin{thebibliography}{130}
	
	
	\bibitem{Gell-Mann:1964ewy}
	M.~Gell-Mann,
	A Schematic Model of Baryons and Mesons,
	Phys. Lett. \textbf{8}, 214-215 (1964).
	
	\bibitem{Zweig:1964ruk}
	G.~Zweig,
	An SU(3) model for strong interaction symmetry and its breaking. Version 1,
	CERN-TH-401.
	
	\bibitem{Zweig:1964jf}
	G.~Zweig,
	An SU(3) model for strong interaction symmetry and its breaking. Version 2,
	CERN-TH-412.
	
	\bibitem{Belle:2003nnu}
	S.~K.~Choi \textit{et al.} [Belle],
	Observation of a narrow charmonium-like state in exclusive $B^\pm \to K^\pm \pi^+ \pi^- J/\psi$ decays,
	Phys. Rev. Lett. \textbf{91}, 262001 (2003).
	
	\bibitem{ParticleDataGroup:2022pth}
	S. Navas \textit{et al.} [Particle Data Group],
	Review of Particle Physics,
	Phys. Rev. D 110, 030001 (2024).
	
	\bibitem{LHCb:2021auc}
	R.~Aaij \textit{et al.} [LHCb],
	Study of the doubly charmed tetraquark $T_{cc}^{+}$,
	Nature Commun. \textbf{13}, 3351 (2022).
	
	\bibitem{LHCb:2021vvq}
	R.~Aaij \textit{et al.} [LHCb],
	Observation of an exotic narrow doubly charmed tetraquark,
	Nature Phys. \textbf{18}, 751-754 (2022).
	
	\bibitem{LHCb:2022sfr}
	R.~Aaij \textit{et al.} [LHCb],
	First Observation of a Doubly Charged Tetraquark and Its Neutral Partner,
	Phys. Rev. Lett. \textbf{131}, 041902 (2023).
	
	\bibitem{LHCb:2020bls}
	R.~Aaij \textit{et al.} [LHCb],
	A model-independent study of resonant structure in $B^+\to D^+D^-K^+$ decays,
	Phys. Rev. Lett. \textbf{125}, 242001 (2020).
	
	\bibitem{LHCb:2015yax}
	R.~Aaij \textit{et al.} [LHCb],
	Observation of $J/\psi p$ Resonances Consistent with Pentaquark States in $\Lambda_b^0 \to J/\psi K^- p$ Decays,
	Phys. Rev. Lett. \textbf{115}, 072001 (2015).
	
	\bibitem{LHCb:2016ztz}
	R.~Aaij \textit{et al.} [LHCb],
	Model-independent evidence for $J/\psi p$ contributions to $\Lambda_b^0\to J/\psi p K^-$ decays,
	Phys. Rev. Lett. \textbf{117}, 082002 (2016).
	
	
	\bibitem{LHCb:2019kea}
	R.~Aaij \textit{et al.} [LHCb],
	Observation of a narrow pentaquark state, $P_c(4312)^+$, and of two-peak structure of the $P_c(4450)^+$,
	Phys. Rev. Lett. \textbf{122}, 222001 (2019).
	
	\bibitem{LHCb:2022ogu}
	R.~Aaij \textit{et al.} [LHCb],
	Observation of a $J/\psi \Lambda$ Resonance Consistent with a Strange Pentaquark Candidate in $B^-\to J/\psi\Lambda \bar{p}$ Decays,
	Phys. Rev. Lett. \textbf{131}, 031901 (2023).
	
	\bibitem{Dong:2021bvy}
	X.~K.~Dong, F.~K.~Guo and B.~S.~Zou,
	A survey of heavy\textendash{}heavy hadronic molecules,
	Commun. Theor. Phys. \textbf{73}, 125201 (2021).
	
	\bibitem{Guo:2017jvc}
	F.~K.~Guo, C.~Hanhart, U.~G.~Mei\ss{}ner, Q.~Wang, Q.~Zhao and B.~S.~Zou,
	Hadronic molecules,
	Rev. Mod. Phys. \textbf{90}, 015004 (2018)
	[erratum: Rev. Mod. Phys. \textbf{94}, no.2, 029901 (2022)].
	
	\bibitem{Liu:2019zoy}
	Y.~R.~Liu, H.~X.~Chen, W.~Chen, X.~Liu and S.~L.~Zhu,
	Pentaquark and Tetraquark states,
	Prog. Part. Nucl. Phys. \textbf{107}, 237-320 (2019).
	
	\bibitem{Chen:2016qju}
	H.~X.~Chen, W.~Chen, X.~Liu and S.~L.~Zhu,
	The hidden-charm pentaquark and tetraquark states,
	Phys. Rept. \textbf{639}, 1-121 (2016).
	
	\bibitem{Brambilla:2015rqa}
	N.~Brambilla, G.~Krein, J.~Tarr\'us Castell\`a and A.~Vairo,
	Long-range properties of $1S$ bottomonium states,
	Phys. Rev. D \textbf{93}, no.5, 054002 (2016).
	
	\bibitem{Dong:2022rwr}
	X.~K.~Dong, F.~K.~Guo, A.~Nefediev and J.~T.~Castell\`a,
	Chromopolarizabilities of fully heavy baryons,
	Phys. Rev. D \textbf{107}, no.3, 034020 (2023).
	
	\bibitem{Dong:2021lkh}
	X.~K.~Dong, V.~Baru, F.~K.~Guo, C.~Hanhart, A.~Nefediev and B.~S.~Zou,
	Is the existence of a $J/\psi J/\psi$ bound state plausible?,
	Sci. Bull. \textbf{66}, no.24, 2462-2470 (2021).
	
	\bibitem{Gong:2020bmg}
	C.~Gong, M.~C.~Du, Q.~Zhao, X.~H.~Zhong and B.~Zhou,
	Nature of X(6900) and its production mechanism at LHCb,
	Phys. Lett. B \textbf{824}, 136794 (2022).
	
	\bibitem{Liu:2023gla}
	W.~Y.~Liu and H.~X.~Chen,
	Fully-heavy hadronic molecules $B_c^{(*)+} B_c^{(*)-}$ bound by fully-heavy mesons,
	[arXiv:2312.11212 [hep-ph]].
	
	\bibitem{Liu:2024pio}
	W.~Y.~Liu and H.~X.~Chen,
	Hadronic molecules with four charm or beauty quarks,
	[arXiv:2405.14404 [hep-ph]].
	
	\bibitem{Liu:2021pdu}
	M.~Z.~Liu and L.~S.~Geng,
	Prediction of an $\Omega_{bbb}\Omega_{bbb}$ Dibaryon in the Extended One-Boson Exchange Model,
	Chin. Phys. Lett. \textbf{38}, no.10, 101201 (2021).
	
	\bibitem{Lyu:2021qsh}
	Y.~Lyu, H.~Tong, T.~Sugiura, S.~Aoki, T.~Doi, T.~Hatsuda, J.~Meng and T.~Miyamoto,
	Dibaryon with Highest Charm Number near Unitarity from Lattice QCD,
	Phys. Rev. Lett. \textbf{127}, no.7, 072003 (2021).
	
	\bibitem{Mathur:2022ovu}
	N.~Mathur, M.~Padmanath and D.~Chakraborty,
	Strongly Bound Dibaryon with Maximal Beauty Flavor from Lattice QCD,
	Phys. Rev. Lett. \textbf{130}, no.11, 111901 (2023).
	
	\bibitem{LHCexp}
	R.~Aaij \textit{et al.} [LHCb],
	Observation of structure in the $J /\psi$-pair mass spectrum,
	Sci. Bull. \textbf{65}, 1983-1993 (2020).
	
	\bibitem{ATLASexp}
	G.~Aad \textit{et al.} [ATLAS],
	Observation of an Excess of Dicharmonium Events in the Four-Muon Final State with the ATLAS Detector,
	Phys. Rev. Lett. \textbf{131}, 151902 (2023).
	
	\bibitem{CMSexp}
	A.~Hayrapetyan \textit{et al.} [CMS],
	Observation of new structure in the J/$\psi$J/$\psi$ mass spectrum in proton-proton collisions at $\sqrt{s}$ = 13 TeV,
	[arXiv:2306.07164 [hep-ex]].
	
	\bibitem{Chao:2020dml}
	K.~T.~Chao and S.~L.~Zhu,
	The possible tetraquark states $cc \bar c \bar c$ observed by the LHCb experiment,
	Sci. Bull. \textbf{65}, 1952-1953 (2020).
	
	\bibitem{liu:2020eha}
	M.~S.~liu, F.~X.~Liu, X.~H.~Zhong and Q.~Zhao,
	Full-heavy tetraquark states and their evidences in the LHCb di-$J/\psi$ spectrum,
	Phys. Rev. D \textbf{109}, 076017 (2024).
	
	\bibitem{Liu:2021rtn}
	F.~X.~Liu, M.~S.~Liu, X.~H.~Zhong and Q.~Zhao,
	Higher mass spectra of the fully-charmed and fully-bottom tetraquarks,
	Phys. Rev. D \textbf{104}, 116029 (2021).
	
\bibitem{Bedolla:2019zwg}
M.~A.~Bedolla, J.~Ferretti, C.~D.~Roberts and E.~Santopinto,
Spectrum of fully-heavy tetraquarks from a diquark+antidiquark perspective,
Eur. Phys. J. C \textbf{80}, 1004 (2020).

	\bibitem{Lu:2020cns}
	Q.~F.~L\"u, D.~Y.~Chen and Y.~B.~Dong,
	Masses of fully heavy tetraquarks $QQ {\bar{Q}} {\bar{Q}}$ in an extended relativized quark model,
	Eur. Phys. J. C \textbf{80}, no.9, 871 (2020).
	
	\bibitem{Wang:2021kfv}
	G.~J.~Wang, L.~Meng, M.~Oka and S.~L.~Zhu,
	Higher fully charmed tetraquarks: Radial excitations and P-wave states,
	Phys. Rev. D \textbf{104}, no.3, 036016 (2021).
	
\bibitem{Zhao:2020jvl}
Z.~Zhao, K.~Xu, A.~Kaewsnod, X.~Liu, A.~Limphirat and Y.~Yan,
Study of charmoniumlike and fully-charm tetraquark spectroscopy,
Phys. Rev. D \textbf{103}, no.11, 116027 (2021).

\bibitem{Zhang:2022qtp}
J.~Zhang, J.~B.~Wang, G.~Li, C.~S.~An, C.~R.~Deng and J.~J.~Xie,
Spectrum of the S-wave fully-heavy tetraquark states,
Eur. Phys. J. C \textbf{82}, 1126 (2022).

\bibitem{Yu:2022lak}
G.~L.~Yu, Z.~Y.~Li, Z.~G.~Wang, J.~Lu and M.~Yan,
The S- and P-wave fully charmed tetraquark states and their radial excitations,
Eur. Phys. J. C \textbf{83}, 416 (2023).

	\bibitem{Badalian:2023rpd}
	A.~M.~Badalian,
	The ${X}$(6550), ${X}$(6900), ${X}$(7280) Resonances as the ${nS}$, ${cc\bar{c}\bar{c}}$ States,
	Phys. Atom. Nucl. \textbf{86}, no.5, 701-708 (2023).
	
	\bibitem{Wu:2024euj}
	W.~L.~Wu, Y.~K.~Chen, L.~Meng and S.~L.~Zhu,
	Benchmark calculations of fully heavy compact and molecular tetraquark states,
	Phys. Rev. D \textbf{109}, no.5, 054034 (2024).
	
\bibitem{Chen:2022asf}
H.~X.~Chen, W.~Chen, X.~Liu, Y.~R.~Liu and S.~L.~Zhu,
An updated review of the new hadron states,
Rept. Prog. Phys. \textbf{86}, no.2, 026201 (2023).

	\bibitem{Zhang:2020vpz}
	J.~R.~Zhang,
	Fully-heavy pentaquark states,
	Phys. Rev. D \textbf{103}, 074016 (2021).
	
	\bibitem{Wang:2021xao}
	Z.~G.~Wang,
	Analysis of the fully-heavy pentaquark states via the QCD sum rules,
	Nucl. Phys. B \textbf{973}, 115579 (2021).

\bibitem{Azizi:2024ito}
K.~Azizi, Y.~Sarac and H.~Sundu,
Investigation of full-charm and full-bottom pentaquark states,
Eur. Phys. J. C \textbf{84}, no.7, 695 (2024).
	
	\bibitem{An:2020jix}
	H.~T.~An, K.~Chen, Z.~W.~Liu and X.~Liu,
	Fully heavy pentaquarks,
	Phys. Rev. D \textbf{103}, 074006 (2021).
	
	\bibitem{Zhang:2023hmg}
	W.~X.~Zhang, H.~T.~An and D.~Jia,
	Masses and magnetic moments of exotic fully heavy pentaquarks,
	Eur. Phys. J. C \textbf{83}, 727 (2023).
	
\bibitem{Rashmi:2024ako}
Rashmi and A.~Upadhyay,
Spectroscopic Analysis of Fully Heavy Pentaquarks,
[arXiv:2410.00633 [hep-ph]].

	\bibitem{An:2022fvs}
	H.~T.~An, S.~Q.~Luo, Z.~W.~Liu and X.~Liu,
	Fully heavy pentaquark states in constituent quark model,
	Phys. Rev. D \textbf{105}, 074032 (2022).

	\bibitem{Yan:2021glh}
	Y.~Yan, Y.~Wu, X.~Hu, H.~Huang and J.~Ping,
	Fully heavy pentaquarks in quark models,
	Phys. Rev. D \textbf{105}, 014027 (2022).
	
	\bibitem{Yan:2023kqf}
	Y.~Yan, Y.~Wu, H.~Huang, J.~Ping and X.~Zhu,
	Prediction of charmed-bottom pentaquarks in quark model,
	Eur. Phys. J. C \textbf{83}, 610 (2023).
	
	\bibitem{Yang:2022bfu}
	G.~Yang, J.~Ping and J.~Segovia,
	Fully charm and bottom pentaquarks in a lattice-QCD inspired quark model,
	Phys. Rev. D \textbf{106}, 014005 (2022).
	
	
\bibitem{Gordillo:2024blx}
M.~C.~Gordillo, J.~Segovia and J.~M.~Alcaraz-Pelegrina,
A diffusion Monte Carlo calculation of fully heavy pentaquarks,
[arXiv:2409.04130 [hep-ph]].

	\bibitem{Gordillo:2023tnz}
	M.~C.~Gordillo and J.~M.~Alcaraz-Pelegrina,
	Asymptotic mass limit of large fully heavy compact multiquarks,
	Phys. Rev. D \textbf{108}, 054027 (2023).
	
	\bibitem{Eichten:1978tg}
	E.~Eichten, K.~Gottfried, T.~Kinoshita, K.~D.~Lane and T.~M.~Yan,
	Charmonium: The Model,
	Phys. Rev. D \textbf{17}, 3090 (1978)
	[erratum: Phys. Rev. D \textbf{21}, 313 (1980)].
	
	\bibitem{Li:2019qsg}
	Q.~Li, M.~S.~Liu, Q.~F.~L\"u, L.~C.~Gui and X.~H.~Zhong,
	Canonical interpretation of $Y(10750)$ and $\Upsilon(10860)$ in the $\Upsilon$ family,
	Eur. Phys. J. C \textbf{80}, 59 (2020).
	
	\bibitem{Deng:2016stx}
	W.~J.~Deng, H.~Liu, L.~C.~Gui and X.~H.~Zhong,
	Charmonium spectrum and their electromagnetic transitions with higher multipole contributions,
	Phys. Rev. D \textbf{95}, 034026 (2017).
	
	\bibitem{Li:2019tbn}
	Q.~Li, M.~S.~Liu, L.~S.~Lu, Q.~F.~L\"u, L.~C.~Gui and X.~H.~Zhong,
	Excited bottom-charmed mesons in a nonrelativistic quark model,
	Phys. Rev. D \textbf{99}, 096020 (2019).
	
	\bibitem{Liu:2019vtx}
	M.~S.~Liu, Q.~F.~L\"u and X.~H.~Zhong,
	Triply charmed and bottom baryons in a constituent quark model,
	Phys. Rev. D \textbf{101}, 074031 (2020).
	
	\bibitem{Liu:2019zuc}
	M.~S.~Liu, Q.~F.~L\"u, X.~H.~Zhong and Q.~Zhao,
	All-heavy tetraquarks,
	Phys. Rev. D \textbf{100}, 016006 (2019).
	
	\bibitem{Varga:1995dm}
	K.~Varga and Y.~Suzuki,
	Precise Solution of Few Body Problems with Stochastic Variational Method on Correlated Gaussian Basis,
	Phys. Rev. C \textbf{52}, 2885-2905 (1995).
	
	\bibitem{Mitroy:2013eom}
	J.~Mitroy, S.~Bubin, W.~Horiuchi, Y.~Suzuki, L.~Adamowicz, W.~Cencek, K.~Szalewicz, J.~Komasa, D.~Blume and K.~Varga,
	Theory and application of explicitly correlated Gaussians,
	Rev. Mod. Phys. \textbf{85}, 693-749 (2013).
	
	\bibitem{Liu:2022hbk}
	F.~X.~Liu, R.~H.~Ni, X.~H.~Zhong and Q.~Zhao,
	Charmed-strange tetraquarks and their decays in the potential quark model,
	Phys. Rev. D \textbf{107}, 096020 (2023).
	
\bibitem{Barnes:1991em}
T.~Barnes and E.~S.~Swanson,
A Diagrammatic approach to meson meson scattering in the nonrelativistic quark potential model,
Phys. Rev. D \textbf{46}, 131-159 (1992).

\bibitem{Barnes:2000hu}
T.~Barnes, N.~Black and E.~S.~Swanson,
Meson meson scattering in the quark model: Spin dependence and exotic channels,
Phys. Rev. C \textbf{63}, 025204 (2001).

\bibitem{Wang:2019spc}
G.~J.~Wang, L.~Y.~Xiao, R.~Chen, X.~H.~Liu, X.~Liu and S.~L.~Zhu,
Probing hidden-charm decay properties of $P_c$ states in a molecular scenario,
Phys. Rev. D \textbf{102}, no.3, 036012 (2020).

\bibitem{Xiao:2019spy}
L.~Y.~Xiao, G.~J.~Wang and S.~L.~Zhu,
Hidden-charm strong decays of the $Z_c$ states,
Phys. Rev. D \textbf{101}, no.5, 054001 (2020).

\bibitem{Wang:2020prk}
G.~J.~Wang, L.~Meng, L.~Y.~Xiao, M.~Oka and S.~L.~Zhu,
Mass spectrum and strong decays of tetraquark ${\bar{c}}{\bar{s}} qq$ states,
Eur. Phys. J. C \textbf{81}, no.2, 188 (2021).

\bibitem{Han:2022fup}
S.~Han and L.~Y.~Xiao,
Aspects of $Z_{cs}(3985)$ and $Z_{cs}(4000)$,
Phys. Rev. D \textbf{105}, no.5, 054008 (2022).

\bibitem{Liu:2024fnh}
F.~X.~Liu, R.~H.~Ni, X.~H.~Zhong and Q.~Zhao,
Hidden and double charm-strange tetraquarks and their decays in a potential quark model,
[arXiv:2407.19494 [hep-ph]].
	
\bibitem{Yang:2021sue}
X.~D.~Yang, F.~L.~Wang, Z.~W.~Liu and X.~Liu,
Newly observed X(4630): a new charmoniumlike molecule,
Eur. Phys. J. C \textbf{81}, no.9, 807 (2021).

\bibitem{Wang:2021hql}
F.~L.~Wang, X.~D.~Yang, R.~Chen and X.~Liu,
Hidden-charm pentaquarks with triple strangeness due to the $\Omega_{c}^{(*)}\bar{D}_s^{(*)}$ interactions,
Phys. Rev. D \textbf{103}, no.5, 054025 (2021).

\bibitem{Liu:2014eka}
X.~H.~Liu, L.~Ma, L.~P.~Sun, X.~Liu and S.~L.~Zhu,
Resolving the puzzling decay patterns of charged $Z_c$ and $Z_b$ states,
Phys. Rev. D \textbf{90}, no.7, 074020 (2014).

	\bibitem{Park:2017jbn}
	W.~Park, A.~Park, S.~Cho and S.~H.~Lee,
	$P_c(4380)$ in a constituent quark model,
	Phys. Rev. D \textbf{95}, 054027 (2017).
	
	\bibitem{Kaeding:1995vq}
	T.~A.~Kaeding,
	Tables of SU(3) isoscalar factors,
	Atom. Data Nucl. Data Tabl. \textbf{61}, 233-288 (1995).
	
	\bibitem{Stancu:1999qr}
	F.~Stancu and S.~Pepin,
	Isoscalar factors of the permutation group,
	Few Body Syst. \textbf{26}, 113-133 (1999).
	
	\bibitem{Hiyama:2003cu}
	E.~Hiyama, Y.~Kino and M.~Kamimura,
	Gaussian expansion method for few-body systems,
	Prog. Part. Nucl. Phys. \textbf{51}, 223-307 (2003).
	
	\bibitem{Brown:2014ena}
	Z.~S.~Brown, W.~Detmold, S.~Meinel and K.~Orginos,
	Charmed bottom baryon spectroscopy from lattice QCD,
	Phys. Rev. D \textbf{90}, 094507 (2014).

	
\end{thebibliography}

\end{document}